%% file: paper.tex
\documentclass[sigplan,screen]{acmart}

\setcopyright{rightsretained}
\acmPrice{}
\acmDOI{10.1145/3385412.3385963}
\acmYear{2020}
\copyrightyear{2020}
\acmSubmissionID{pldi20main-p24-p}
\acmISBN{978-1-4503-7613-6/20/06}
\acmConference[PLDI '20]{Proceedings of the 41st ACM SIGPLAN International Conference on Programming Language Design and Implementation}{June 15--20, 2020}{London, UK}
\acmBooktitle{Proceedings of the 41st ACM SIGPLAN International Conference on Programming Language Design and Implementation (PLDI '20), June 15--20, 2020, London, UK}

\startPage{1}

\usepackage{booktabs}
\usepackage{graphicx}
\usepackage{amsmath}
\usepackage{amssymb}
\usepackage{algorithmic}
\usepackage{algorithm}
\usepackage{subcaption}
\usepackage{url}
\usepackage{siunitx}
\usepackage{xspace}
\usepackage{listings}
\usepackage{microtype}
\usepackage{multirow}
\usepackage{multicol}
\usepackage{tabularx}
\usepackage{mathtools}
\usepackage{enumitem}
\usepackage{adjustbox}
\usepackage{wrapfig}
\usepackage{syntax}
\usepackage{color}
\usepackage{colortbl}
\usepackage{placeins}
\usepackage{hyperref}

\DeclareMathOperator{\assign}{=}
\DeclareMathOperator{\map}{map}
\newcommand{\reduce}[1]{\mathrel{#1}=}
\newcommand{\iteration}[1]{\forall_{#1} \;}
\newcommand{\titeration}[1]{\forall_{#1}}

\DeclareMathOperator{\where}{\textbf{where}}

\definecolor{todocolor}{rgb}{0.8,0,0}
\definecolor{keywordcolor}{rgb}{0.5,0,0.5}
\definecolor{nonsymmcolor}{gray}{0.9}

\newcommand{\figref}[1]{Figure~\ref{fig:#1}}
\newcommand{\figsref}[2]{Figures~\ref{fig:#1} and~\ref{fig:#2}}

\newcommand{\secref}[1]{Section~\ref{sec:#1}}

\newcommand{\secsrangeref}[2]{Sections~\ref{sec:#1} through~\ref{sec:#2}}

\newcommand{\tabref}[1]{Table~\ref{tab:#1}}
\newcommand{\HIDE}[1]{}

\ifdefined\notodo
\newcommand{\TODO}[1]{}
\else
\newcommand{\TODO}[1]{{\color{todocolor}#1}}
\fi

\newcolumntype{R}[2]{%
  >{\adjustbox{angle=#1,lap=\width-(#2)}\bgroup}%
  l%
  <{\egroup}%
}

\newcolumntype{R}{>{\raggedleft\arraybackslash}X}

\definecolor{textgray}{gray}{0.4}
\lstdefinestyle{levelfunc}{
  language=C++,
  mathescape,
  frame=none,
  aboveskip=\medskipamount,
  belowskip=\medskipamount,
  columns=flexible,
  basicstyle=\fontsize{9}{9}\ttfamily,
  keywordstyle=\color{keywordcolor},
  commentstyle=\color{gray},
  showstringspaces=false,
}
\newcommand\code[1]{\lstinline[columns=fullflexible, mathescape, basicstyle=\fontsize{9}{9}\ttfamily]|#1|}
\newcommand\smallcode[1]{\lstinline[columns=fullflexible, mathescape, basicstyle=\fontsize{8}{8}\ttfamily]|#1|}
\newcommand\query[1]{\lstinline[language=sql, columns=fullflexible, mathescape, morekeywords={id} , basicstyle=\fontsize{9}{9}\ttfamily]|#1|}

\SetEnumitemKey{twocol}{
  before=\raggedcolumns\setlength{\multicolsep}{\topsep}\begin{multicols}{2},
  after=\end{multicols}
}

\begin{document}

\title[Automatic Generation of Efficient Sparse Tensor Format Conversion Routines]{Automatic Generation of Efficient \\ Sparse Tensor Format Conversion Routines}

\author{Stephen Chou}
\affiliation{
  \institution{MIT CSAIL}
  \streetaddress{32-G778, 32 Vassar Street}
  \city{Cambridge}
  \state{MA}
  \postcode{02139}
  \country{USA}
}
\email{s3chou@csail.mit.edu}

\author{Fredrik Kjolstad}
\affiliation{
  \institution{Stanford University}
  \streetaddress{440, 353 Serra Mall}
  \city{Stanford}
  \state{CA}
  \postcode{94305}
  \country{USA}
}
\email{kjolstad@cs.stanford.edu}

\author{Saman Amarasinghe}
\affiliation{
  \institution{MIT CSAIL}
  \streetaddress{32-G744, 32 Vassar Street}
  \city{Cambridge}
  \state{MA}
  \postcode{02139}
  \country{USA}
}
\email{saman@csail.mit.edu}

\begin{abstract}

  This paper shows how to generate code that efficiently converts sparse tensors between disparate storage formats (data layouts) such as CSR, DIA, ELL, and many others.
  We decompose sparse tensor conversion into three logical phases: coordinate remapping, analysis, and assembly.
  We then develop a language that precisely describes how different formats group together and order a tensor's nonzeros in memory.
  This lets a compiler emit code that performs complex remappings of nonzeros when converting between formats.
  We also develop a query language that can extract statistics about sparse tensors, and we show how to emit efficient analysis code that computes such queries. 
  Finally, we define an abstract interface that captures how data structures for storing a tensor can be efficiently assembled given specific statistics about the tensor.
  Disparate formats can implement this common interface, thus letting a compiler emit optimized sparse tensor conversion code for arbitrary combinations of many formats without hard-coding for any specific combination.

  Our evaluation shows that the technique generates sparse tensor conversion routines with performance between 1.00 and 2.01$\times$ that of hand-optimized versions in SPARSKIT and Intel MKL, two popular sparse linear algebra libraries.
  And by emitting code that avoids materializing temporaries, which both libraries need for many combinations of source and target formats, our technique outperforms those libraries by 1.78 to 4.01$\times$ for CSC/COO to DIA/ELL conversion.
  
\end{abstract}

\ccsdesc[500]{Software and its engineering~Abstraction, modeling and modularity} 
\ccsdesc[500]{Software and its engineering~Source code generation}               
\ccsdesc[300]{Software and its engineering~Domain specific languages}            
\ccsdesc[300]{Mathematics of computing~Mathematical software performance}        
\keywords{sparse tensor conversion, sparse tensor assembly, sparse tensor algebra, sparse tensor formats, coordinate remapping notation, attribute query language}


\maketitle

\section{Introduction}
\label{sec:introduction}
\input{sections/introduction}

\section{Background}
\label{sec:background}
\input{sections/background.tex}

\section{Overview}
\label{sec:tensor-format-conversion}
\input{sections/tensor-format-conversion}

\section{Coordinate Remapping}
\label{sec:coordinate-remapping}
\input{sections/coordinate-remapping}

\section{Attribute Queries}
\label{sec:attribute-queries}
\input{sections/attribute-queries}

\section{Sparse Tensor Assembly}
\label{sec:tensor-assembly}
\input{sections/assembly-abstraction}

\section{Evaluation}
\label{sec:evaluation}
\input{sections/evaluation}

\section{Related Works}
\label{sec:related-works}
\input{sections/related-works}

\section{Conclusion and Future Work}
\label{sec:conclusions}
\input{sections/conclusions}

\begin{acks}
  We thank Peter Ahrens, Rawn Henry, Ziheng Wang, Michelle Strout, and other anonymous reviewers for their helpful reviews and suggestions.
  This work was supported by the Application Driving Architectures (ADA) Research Center, a JUMP Center co-sponsored by SRC and DARPA; the Toyota Research Institute; the U.S. Department of Energy, Office of Science, Office of Advanced Scientific Computing Research under Award Numbers DE-SC0008923 and DE-SC0018121; the National Science Foundation under Grant No. CCF-1533753; and DARPA under Awards HR0011-18-3-0007 and HR0011-20-9-0017.
  Any opinions, findings, and conclusions or recommendations expressed in this material are those of the authors and do not necessarily reflect the views of the aforementioned funding agencies.
\end{acks}

\balance
\bibliography{paper} 

%



\end{document}

%% file: sections/introduction.tex
Sparse multidimensional arrays (tensors) are suited for representing data in many domains, including data analytics~\cite{anandkumartensor,bader2008}, machine learning~\cite{Rajbhandari2017,park2016faster}, and others.
Countless formats for storing sparse tensors have been developed~\cite{hicoo,Xie2018,bulucc2009,smith2015tensor,csr5,Bell2009,Ashari2014,Saad1989,Baskaran2012,bcsr,fcoo,bader2007,saad2003,csr,ell,dcsc,bicrs,sell,clspmv} to accelerate kernels like sparse matrix-vector multiplication (SpMV), and new formats are constantly being proposed in recent literature.

No format is universally superior in every circumstance, since the ideal format for storing a sparse tensor depends on its structure and sparsity, the operation being performed, and the available hardware.
Applications typically need to perform different operations on the same tensor, and each operation may require the tensor to be stored in a distinct format for optimal performance.
Importing data into a sparse tensor, for instance, can be done efficiently if the tensor is constructed in the COO format~\cite{bader2007} or the DOK format~\cite{dok}, since they support efficient appends or random insertions of new nonzeros.
Computing SpMV with the tensor, however, can be done more than twice as fast if the tensor is stored in CSR~\cite{csr}, which compresses out redundant row coordinates and thereby reduces memory traffic~\cite{chou2018}.
Alternatively, if all of the tensor's nonzeros are clustered along a few dense diagonals, then storing it in DIA~\cite{saad2003} minimizes memory traffic even more while exposing vectorization opportunities, further improving SpMV performance by up to 22\% as a result~\cite{chou2018}.
Thus, to optimize the performance of both data import and compute, an application must convert the tensor from COO (or DOK) to DIA (or CSR).
And in applications like preconditioned solvers and sparse neural network training where a tensor might only be computed with a few times, the conversion must be efficient so that the overhead does not outweigh gains from using an optimized format~\cite{elafrou2017}.

General-purpose sparse linear and tensor algebra libraries like SPARSKIT~\cite{Saad94sparskit} and Intel MKL~\cite{mkl} thus strive to support efficiently converting tensors between as many formats as possible.
SPARSKIT, for instance, supports no fewer than 17 different matrix formats.
With such large numbers ($N$) of supported formats, it becomes impractical to manually implement efficient conversion routines for all $\Theta(N^2)$ combinations of source and target formats.
Instead, hand-optimized libraries typically only support direct conversions to and from some arbitrary canonical format (e.g., CSR with SPARSKIT).
Thus, to convert a tensor from COO to DIA using SPARSKIT (or Intel MKL), an application must first convert the tensor to CSR and then to DIA.
This doubles the number of conversions needed, which is inefficient when converting a tensor even once incurs significant overhead~\cite{Xie2018,chou2018}.
Worse, this approach is not even feasible if an application uses novel formats that are not supported by libraries, such as many of the sparse matrix and tensor formats that have been proposed in recent literature.
The developer must then hand-implement efficient custom conversion routines for each new format, which are typically complicated and tedious to write and optimize.
This motivates a technique that can instead automatically generate such efficient format conversion routines.

\begin{figure}
  \centering
  \includegraphics[width=\columnwidth]{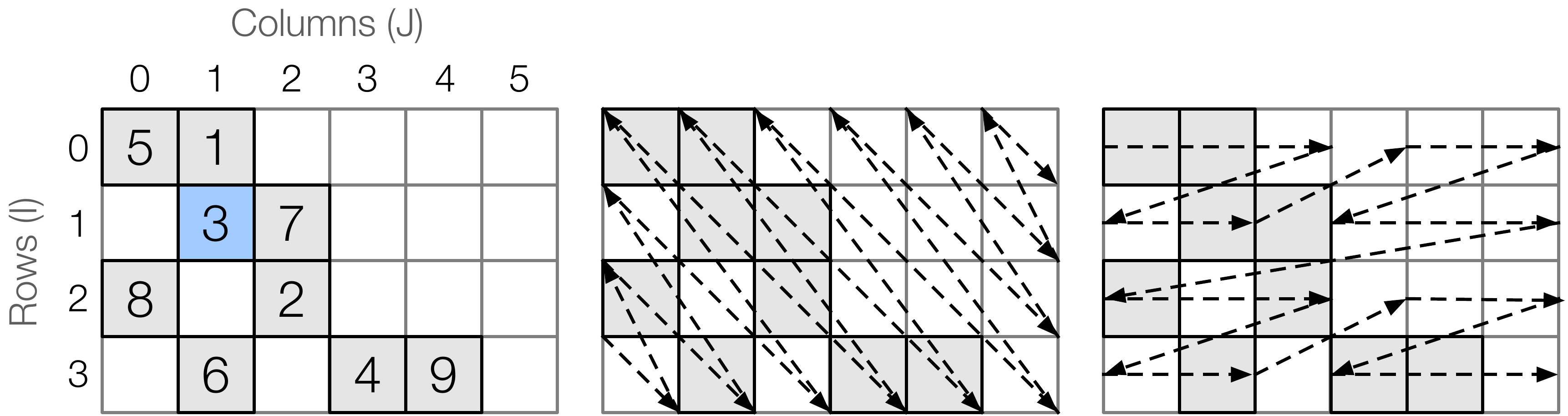}
  \caption {
    A sparse 4$\times$6 tensor (matrix).
    Nonzeros can be ordered in memory not only by rows or by columns, but also by diagonals (center) or even by blocks (right).
   }
  \label{fig:matrix-example}
\end{figure}

Existing sparse tensor algebra compilers such as taco~\cite{kjolstad2017,chou2018,kjolstad2019} are unable to generate such routines for many formats.
Converting a tensor between disparate formats typically entails changing how its nonzeros are grouped and ordered in memory, potentially rearranging nonzeros into complex orders such as by diagonals or by blocks (\figref{matrix-example}), or even by Morton order~\cite{bulucc2009,hicoo}.
Efficient conversion algorithms can often achieve this reordering without explicitly sorting the input tensor by first computing statistics about the tensor.
These statistics are then used to coordinate the movement of nonzeros to the output tensor (in the target format) in such a way that avoids data reshuffles and memory reallocations.
The taco compiler cannot generate such algorithms as it cannot express or reason about reordering nonzeros in non-lexicographic coordinate order.
It also cannot reason about computing and utilizing statistics about the input tensor to coordinate assembly of disparate tensor data structures.

We propose a technique to generate efficient sparse tensor conversion routines for a wide range of disparate formats, building on our recent works on sparse tensor algebra compilation~\cite{kjolstad2017,chou2018,kjolstad2019}.
We decompose a large class of tensor conversion algorithms into three logical phases (\secref{tensor-format-conversion}).
Then, to facilitate generating code for each phase, we develop 
\begin{description}

  \item[coordinate remapping notation,] which describes how different tensor formats group together and order nonzeros in memory (\secref{coordinate-remapping});

  \item[attribute query language,] which describes what statistics about a tensor are needed so that sufficient memory can be reserved for conversion (\secref{attribute-queries}); and a 

  \item[tensor assembly abstract interface,]  which exposes functions that capture how results of attribute queries are used to efficiently assemble many kinds of sparse tensor data structures (\secref{tensor-assembly}).

\end{description}
As we will show, the conciseness of these abstractions makes it easy to provide specifications that describe how to efficiently construct sparse tensors in many formats.
Our technique can then combine these specifications with additional ones proposed by \citeauthor{chou2018}~\shortcite{chou2018}, which specify how to efficiently iterate over sparse tensors in many formats, in order to generate efficient conversion routines for arbitrary combinations of formats.
In this way, users only have to provide one set of specifications for every supported format rather than every combination of source and target formats. 

We have implemented a prototype of our technique in taco.
Our evaluation shows that, for many combinations of source and target formats, our technique generates conversion routines with performance between 1.00 and 2.01$\times$ that of hand-optimized implementations in SPARSKIT and Intel MKL. 
For conversions from CSC/COO to DIA/ELL, our technique emits code that avoid materializing temporaries and that are not implemented in either library, which lets us optimize those conversions by 1.78 to 4.01$\times$ (\secref{evaluation}).

%% file: sections/background.tex
\begin{figure*}
  \begin{minipage}[b]{0.22\linewidth}
    \centering
    \includegraphics[scale=0.25]{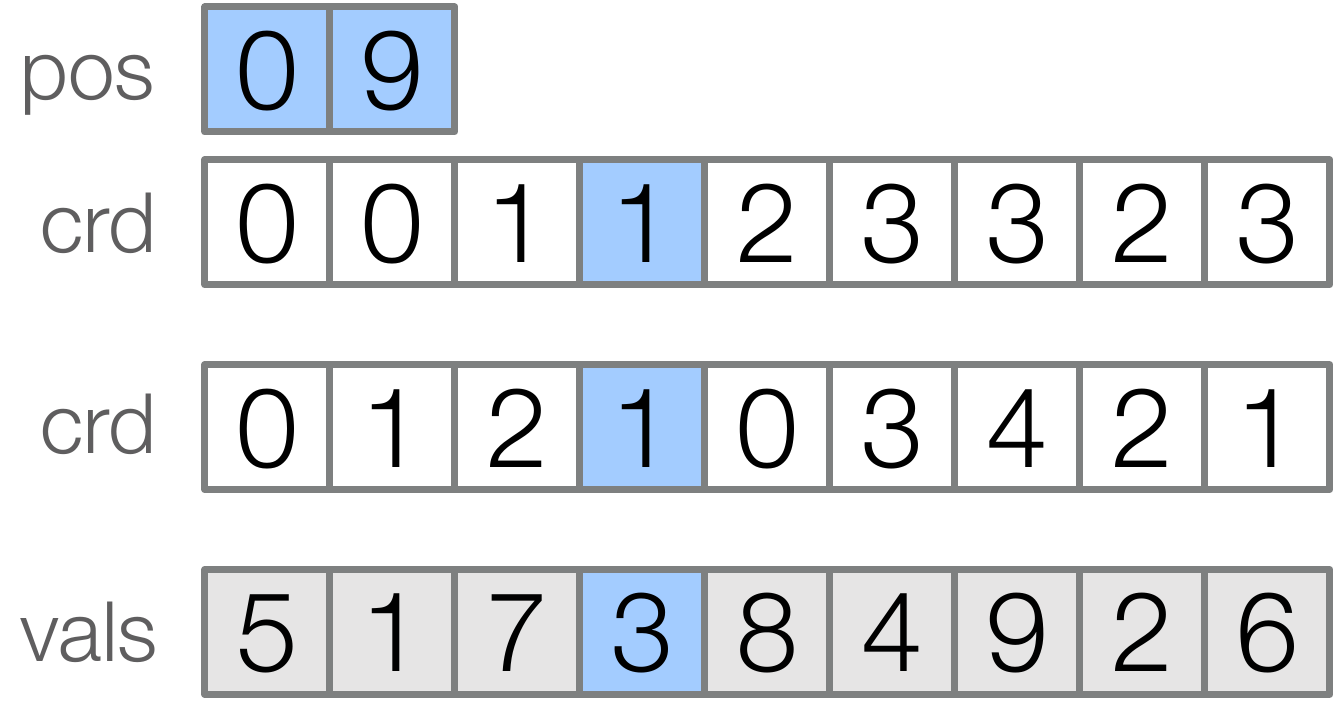}
    \subcaption{
      COO
    }
    \label{fig:matrix-example-coo}
  \end{minipage}
  \begin{minipage}[b]{0.22\linewidth}
    \centering
    \includegraphics[scale=0.25]{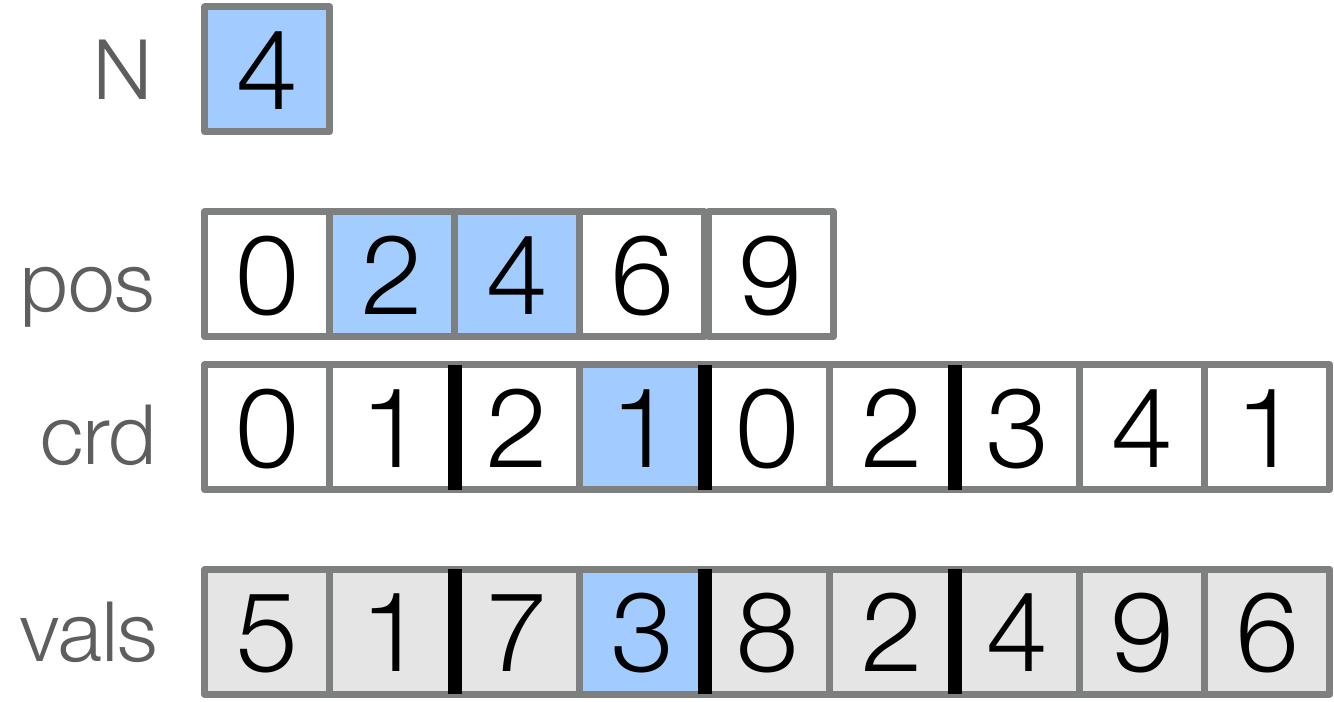}
    \subcaption{
      CSR
    }
    \label{fig:matrix-example-csr}
  \end{minipage}
  \begin{minipage}[b]{0.27\linewidth}
    \centering
    \includegraphics[scale=0.25]{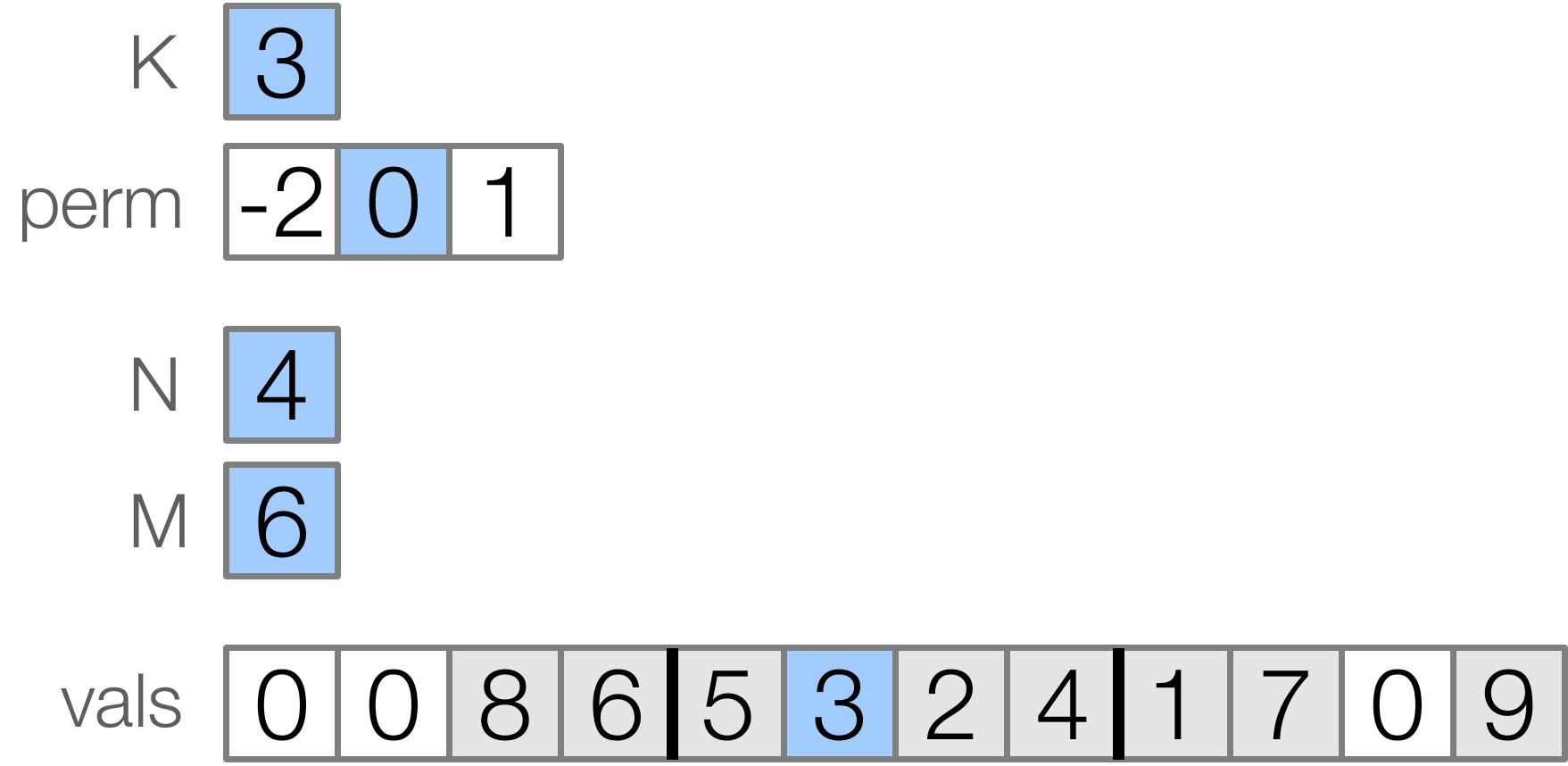}
    \subcaption{
      DIA
    }
    \label{fig:matrix-example-dia}
  \end{minipage}
  \begin{minipage}[b]{0.27\linewidth}
    \centering
    \includegraphics[scale=0.25]{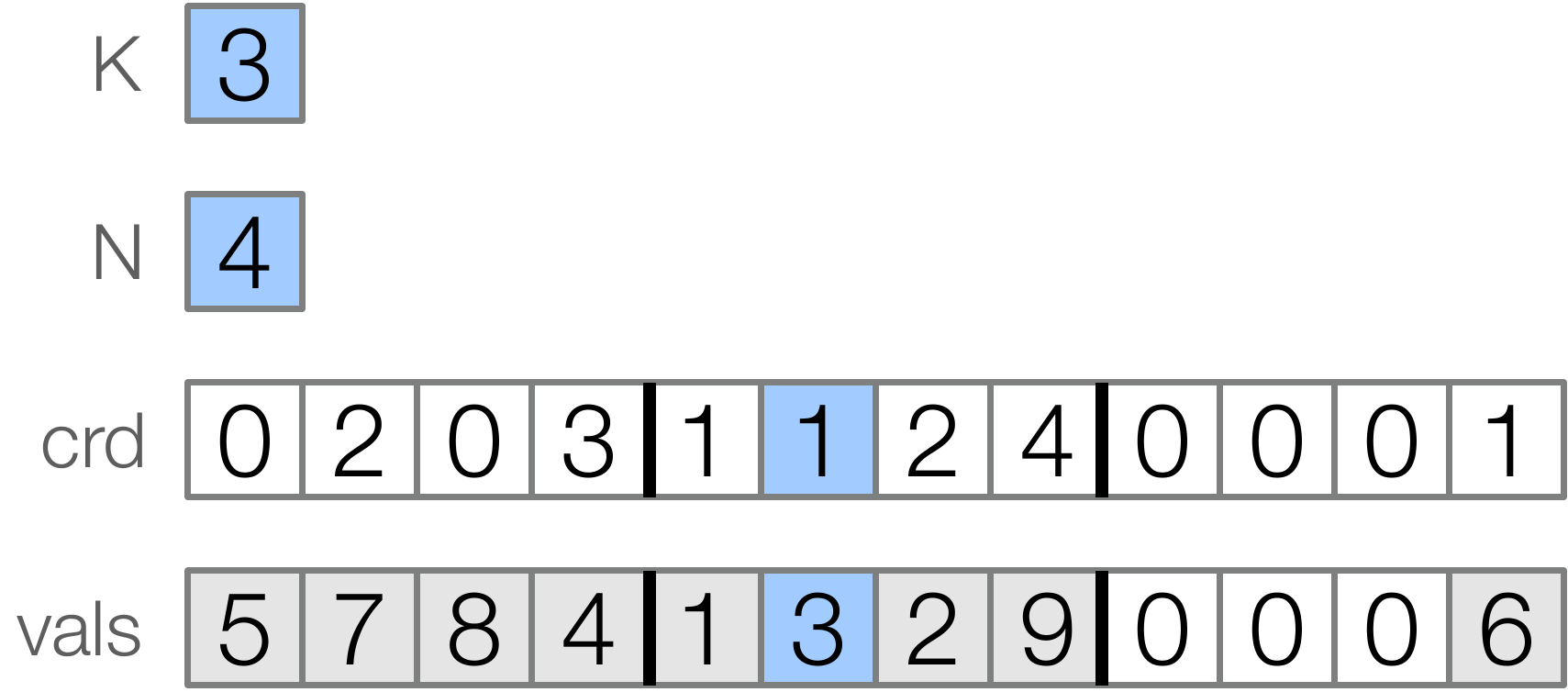}
    \subcaption{
      ELL
    }
    \label{fig:matrix-example-ell}
  \end{minipage}
  \caption{
    The same tensor as shown in \figref{matrix-example}, stored in disparate sparse tensor formats.
  }
  \label{fig:matrix-formats}
\end{figure*}

There exist a wide variety of formats for storing sparse tensors in memory.
\figref{matrix-formats} shows four examples of commonly used sparse tensor formats; for an overview of more formats that have been proposed, we refer readers to \href{https://arxiv.org/pdf/1804.10112.pdf#subsection.2.1}{Section 2.1} in~\cite{chou2018}.
The COO format~\cite{bader2007} represents a sparse tensor as a list of its nonzeros, storing the complete coordinates and value of each nonzero.
COO supports efficiently appending new nonzeros, though it also wastes memory by storing redundant row coordinates.
The CSR format~\cite{csr} compresses out the redundant row coordinates by grouping all nonzeros in the same row together and using a \code{pos} array to map nonzeros to each row.
However, inserting a nonzero at some arbitrary coordinates into CSR is expensive as all nonzeros in subsequent rows must be shifted in memory.
The DIA~\cite{saad2003} format stores nonzeros along the same diagonal together in memory, while the ELL~\cite{ell} format groups together up to one nonzero from each row.
Such orderings of nonzeros expose vectorization opportunities for SpMV~\cite{DAzevedo2005} and can also reduce memory footprint.
However, DIA is only suitable for diagonal and banded matrices, while ELL is only suitable if all rows in the matrix have a similar number ($K$) of nonzeros.
As these examples show, there is no universally ideal format for storing sparse matrices.
The same is true of higher-order sparse tensors, for which even more formats have been---and are continuously being---proposed~\cite{smith2015tensor,hicoo,fcoo,Baskaran2012} that use disparate data structures and ordering schemes to store nonzeros, each with distinct trade-offs.

\begin{figure}
  \centering
  \includegraphics[scale=0.5]{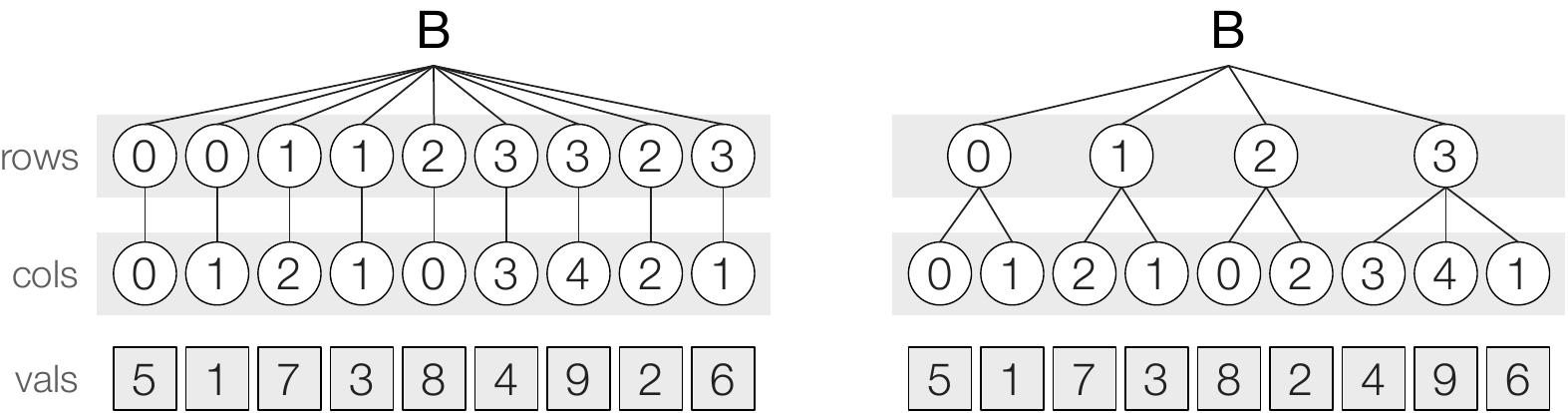}
  \caption {
    Two coordinate hierarchy representations of the same tensor from \figref{matrix-example} in COO (left) and CSR (right).
    Their differing structures reflect how COO and CSR store nonzeros (i.e., whether duplicate row coordinates are stored).
  }
  \label{fig:coordinate-hierarchy-examples}
\end{figure}

\citeauthor{chou2018}~\shortcite{chou2018} describe how tensors stored in disparate formats can be represented as \emph{coordinate hierarchies} that have varying structures but that expose the same abstract interface.
\figref{coordinate-hierarchy-examples} shows examples of coordinate hierarchies that represent a tensor in two different formats. 
Each level in a coordinate hierarchy encodes the nonzeros' coordinates into one dimension.
Edges associate stored components with their containing subtensors.
In \figref{coordinate-hierarchy-examples}, for instance, each column coordinate, which is associated with a nonzero, is connected by an edge to a coordinate that identifies the nonzero's containing row.
Each stored component is represented by a path from the root to a leaf, with coordinates along the path representing the component's coordinates.
We refer readers to \href{https://arxiv.org/pdf/1804.10112.pdf#section.3}{Section 3} in~\cite{chou2018} for more details.

\begin{figure}
  \centering
  \includegraphics[width=\linewidth]{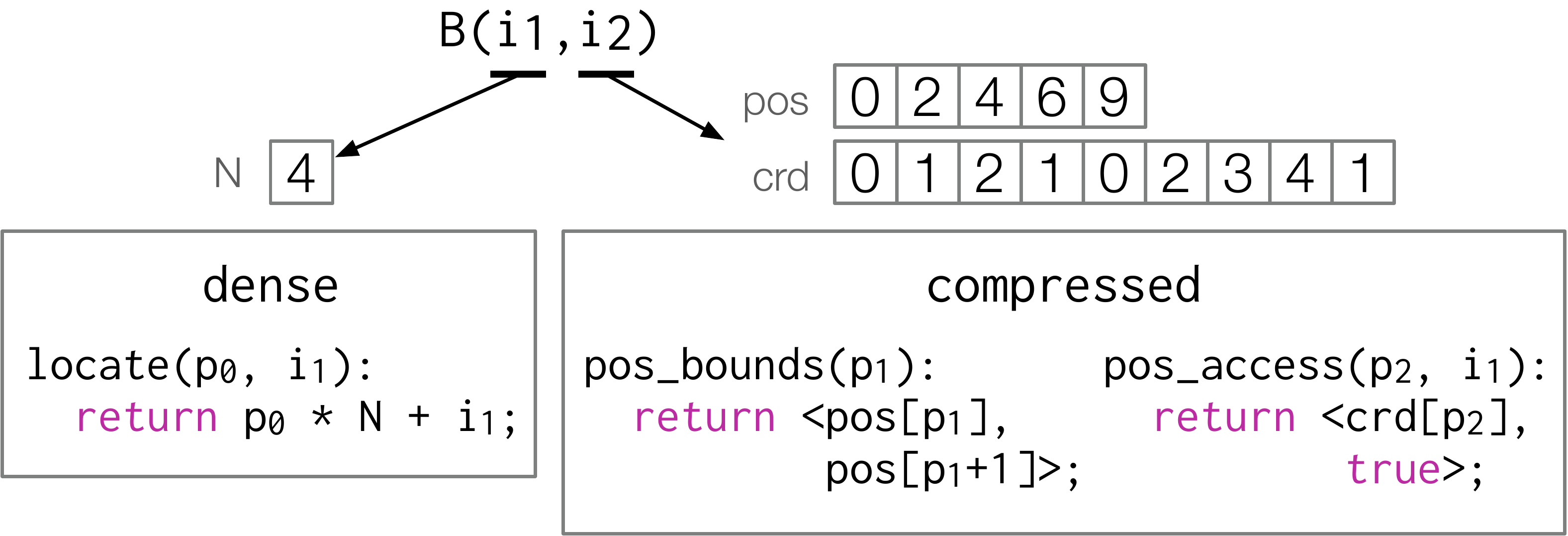}
  \caption {
    Decomposition of CSR into level formats and corresponding level functions that describe how the associated data structures can be efficiently accessed.
  }
  \label{fig:csr-decomposition}
\end{figure}

We can then decompose sparse tensor formats into \emph{level formats} that each stores a coordinate hierarchy level, which represents a tensor dimension.
CSR (\figref{matrix-example-csr}), for instance, can be decomposed into two level formats, \emph{dense} and \emph{compressed}, that store the row and column levels respectively, as \figref{csr-decomposition} shows.
The dense level format implicitly encodes all rows using just a parameter \code{N} to store the dimension's size.
By contrast, the compressed level format uses two arrays, \code{pos} and \code{crd}, to store column coordinates of nonzeros in each row.
All level formats, however, expose the same static interface consisting of \emph{level functions}, which describe how to access a format's data structures, and \emph{properties}, which describe characteristics of the data as stored (e.g., if nonzeros are stored in order).
The level function \code{locate}, for instance, describes how to efficiently random access coordinates that are stored in a level format.
Similarly, \code{pos_bounds} and \code{pos_access} describe how to efficiently iterate over coordinates, with the former specifying how to compute the bounds of iteration and the latter specifying how to access each coordinate.

\begin{figure}
  \centering
  \includegraphics[width=\linewidth]{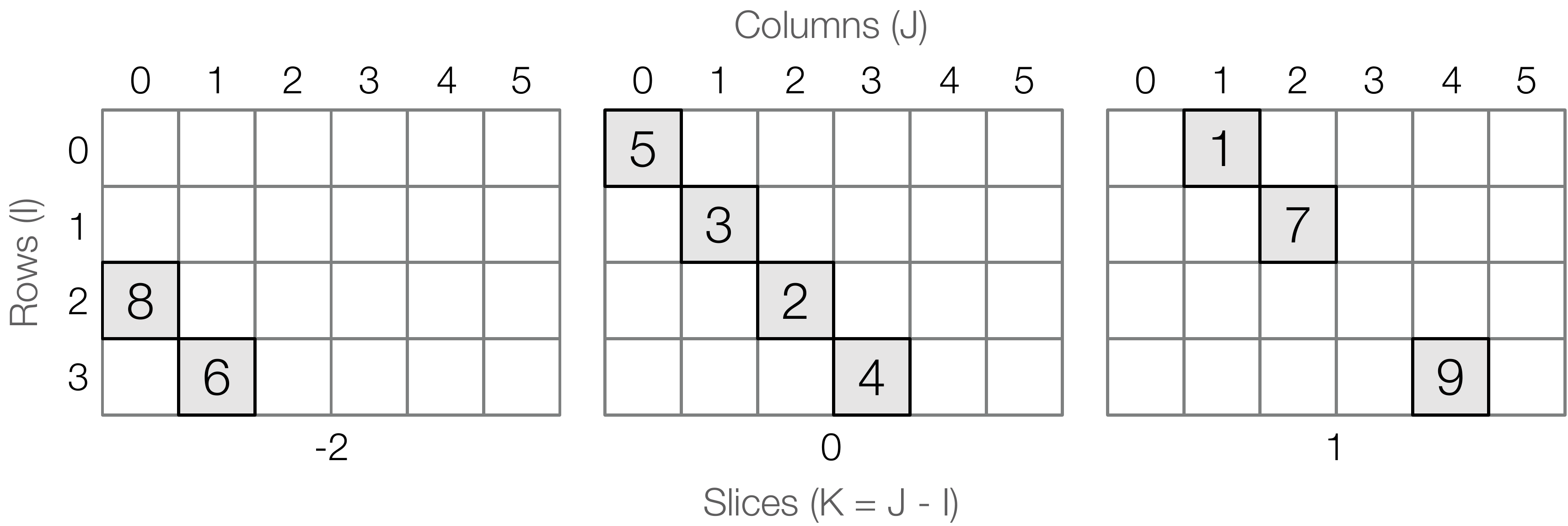}
  \caption{
    The matrix in \figref{matrix-example} can be transformed to a 3rd-order tensor where each slice contains all nonzeros that lie on the same diagonal in the original matrix.
    The lexicographic coordinate ordering of nonzeros in the resulting tensor matches the order in which nonzeros are stored in DIA (\figref{matrix-example-dia}).
    \secref{coordinate-remapping} shows how this transformation is formalized by the coordinate remapping \code{(i,j) ->} \code{(j-i,i,j)}.
  }
  \label{fig:dia-tensorization}
\end{figure}

Structured sparse tensor formats like DIA and ELL, which do not group nonzeros lexicographically by their coordinates, can also be decomposed into level formats by casting them as formats for tensors with additional dimensions.
For example, a DIA matrix can be cast as a 3rd-order tensor where each slice contains only nonzeros that lie on the same diagonal, as shown in \figref{dia-tensorization}.
We can then decompose DIA into three level formats: one that stores the set of nonzero diagonals in a \code{perm} array of size \code{K}, another that encodes the set of rows in each diagonal, and a third that encodes the column coordinates of nonzeros.
Such a decomposition lets a sparse tensor algebra compiler reason about how tensors stored in DIA and similar structured formats can be efficiently iterated, which is crucial for generating fast tensor algebra code.

The coordinate hierarchy abstraction lets a compiler generate efficient code to iterate over sparse tensors in disparate formats by simply emitting code to traverse coordinate hierarchies.
This entails recursively generating a set of nested loops that each iterates over a level in a coordinate hierarchy.
The compiler generates each loop by emitting calls to level functions that describe how to efficiently access the level.
Then, to obtain code that iterates over a tensor in any desired format, the level function calls are simply replaced with the desired format's implementations of those level functions.
This approach lets a compiler generate efficient code for disparate formats without hard-coding for any specific format.
We refer readers to \href{https://arxiv.org/pdf/1804.10112.pdf#subsection.4.5}{Section 4.5} in~\cite{chou2018} for more details.

%% file: sections/tensor-format-conversion.tex
\begin{figure*}[!htb] 
  \begin{minipage}[t]{0.33\linewidth}
    \centering
    \vspace{0pt}
    \includegraphics[width=\linewidth]{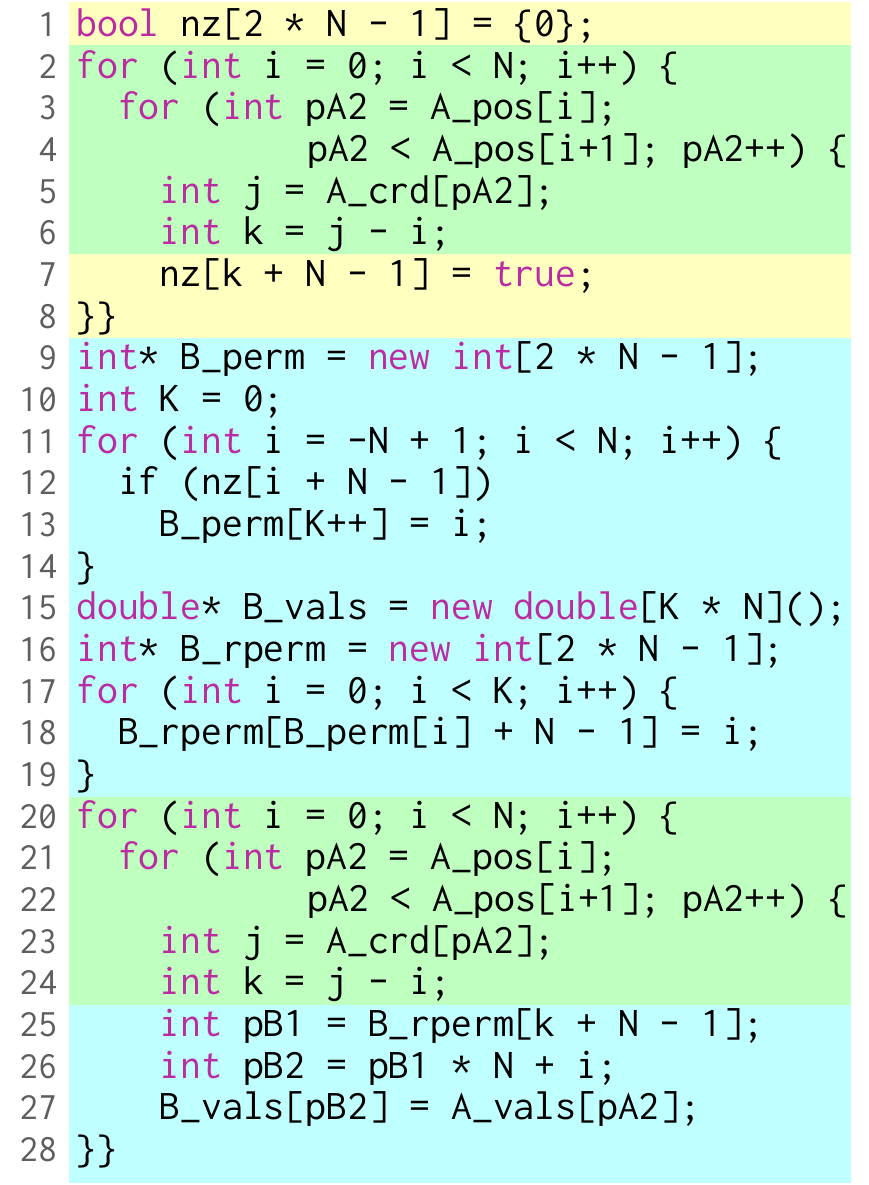}
    \subcaption{
      CSR to DIA
    }
    \label{fig:csrdia-example}
  \end{minipage}
  \hfill
  \begin{minipage}[t]{0.33\linewidth}
    \centering
    \vspace{0pt}
    \includegraphics[width=\linewidth]{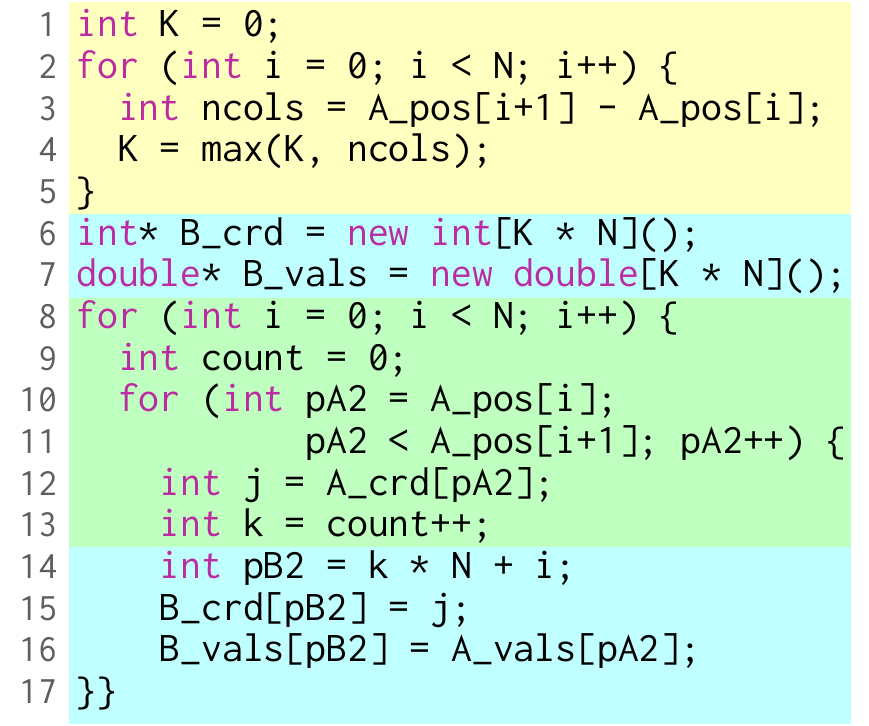}
    \vspace{76.4pt}
    \subcaption{
      CSR to ELL
    }
    \label{fig:csrell-example}
  \end{minipage}
  \hfill
  \begin{minipage}[t]{0.33\linewidth}
    \centering
    \vspace{0pt}
    \includegraphics[width=\linewidth]{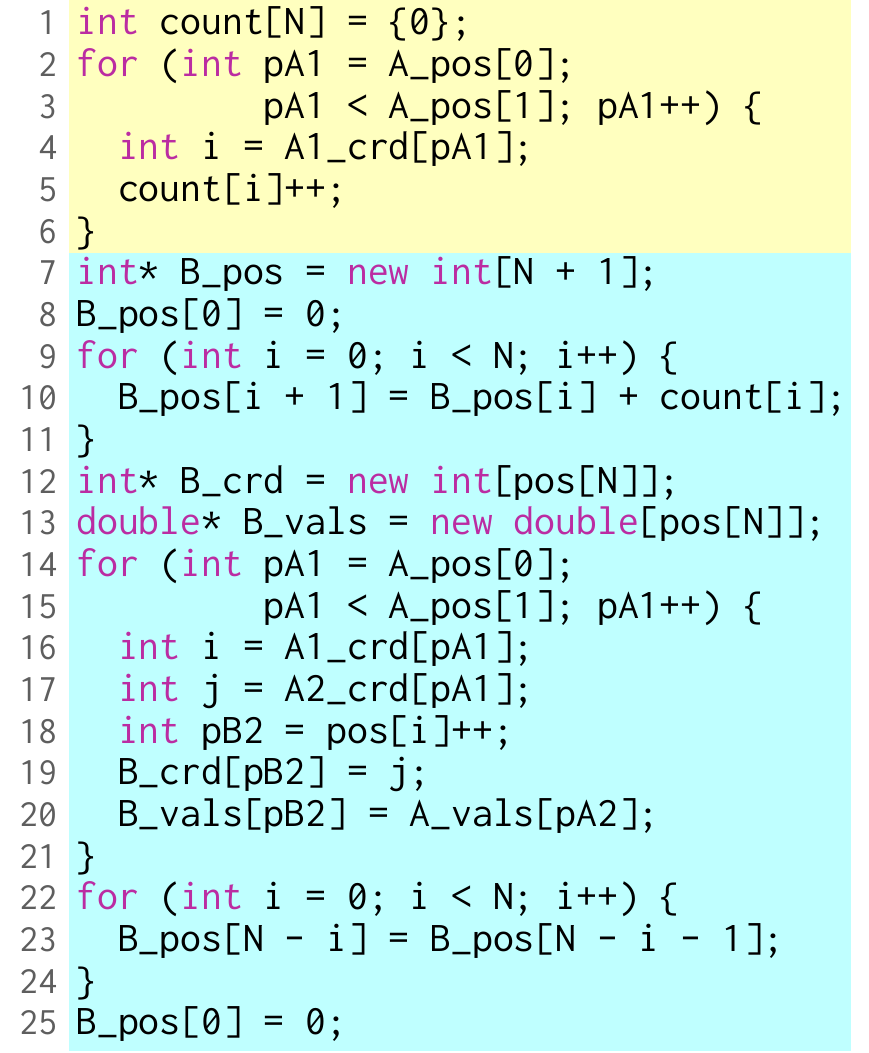}
    \vspace{13.5pt}
    \subcaption{
      COO to CSR
    }
    \label{fig:coocsr-example}
  \end{minipage}
  \caption{
    Code (in C++) to convert sparse tensors between different source and target formats.
    The background colors identify distinct logical phases of format conversion (green for coordinate remapping, yellow for analysis, and blue for assembly).
  }
  \label{fig:conversion-examples}
\end{figure*}

\begin{figure*}
  \centering
  \includegraphics[width=\linewidth]{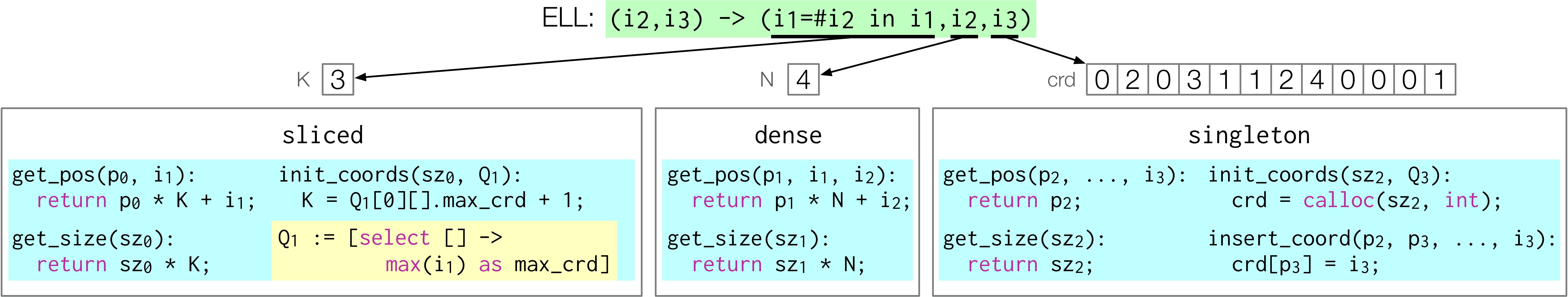}
  \caption{
    Specifications that describe what needs to be performed as part of each logical phase when converting sparse tensors to ELL, which can be decomposed into three level formats: \emph{sliced}, \emph{dense}, and \emph{singleton}.
    The background colors identify the phases being specified, following the same scheme as \figref{conversion-examples}.
  }
  \label{fig:ell-decomposition}
\end{figure*}

\figref{conversion-examples} shows three examples of sparse tensor conversion routines that efficiently convert tensors between different storage formats.
As these examples illustrate, different combinations of source and target formats require vastly dissimilar code.
It turns out, however, that efficient algorithms for converting sparse tensors between a wide range of disparate formats can all be decomposed into three logical phases: coordinate remapping, analysis, and assembly.
\figref{conversion-examples} identifies these phases using different background colors.

The coordinate remapping phase iterates over the input tensor and, for each nonzero, computes new coordinates as functions of its original coordinates.
What additional coordinates are computed depends on the target format. 
For instance, the code in \figref{csrdia-example}, which converts a CSR matrix to DIA, computes a new coordinate \code{k} for each nonzero as the difference between its column and row coordinates (lines 2--6 and 20--24).
Coordinate remapping conceptually transforms (i.e., remaps) the input tensor to a hypersparse higher-order tensor.
As \figref{dia-tensorization} illustrates, the lexicographic coordinate ordering of nonzeros in the remapped tensor reflects how nonzeros are grouped together and ordered in the target format (i.e., DIA), even if the format does not store nonzeros in lexicographic order by their original coordinates.

The analysis phase computes statistics about the input tensor that are later used to determine the amount of memory to allocate for storing nonzeros in the target format.
The exact statistics that are computed also depend on the target format.
\figref{csrdia-example}, for instance, computes the set of all nonzero diagonals in the input matrix (lines 1--8), with distinct diagonals identified by offsets (\code{k}) computed in the coordinate remapping phase.
By contrast, \figref{csrell-example} computes the maximum number of nonzeros in any row of the input matrix (lines 1--5), while \figref{coocsr-example} computes the number of nonzeros in each row of the input matrix (lines 1--6).

Finally, the assembly phase iterates over the input tensor and inserts each nonzero into the output data structures.
Again, where each nonzero is inserted (\code{pB2}) depends on the target format.
\figref{csrdia-example} computes \code{pB2} as a function of each nonzero's row coordinate and its offset \code{k} (as computed in the coordinate remapping phase), in such a way that nonzeros with the same offset are grouped together in the output (lines 25--26).
By contrast, \figref{coocsr-example} simply appends each nonzero to its row's corresponding segment in the \code{crd} array (line 19).

In \secsrangeref{coordinate-remapping}{tensor-assembly}, we describe how our technique generates efficient code to perform coordinate remapping (\secref{coordinate-remapping}), analysis (\secref{attribute-queries}), and assembly (\secref{tensor-assembly}).
For each logical phase, we define a language that captures what needs to be performed for disparate target formats.
\figref{ell-decomposition} demonstrates how these languages can be used to specify what needs to be performed when converting tensors to ELL.
For each new target tensor format, a user must first specify
\begin{itemize}[leftmargin=1.25\parindent]
  \item a coordinate remapping (in green) that, when applied to the input tensor, captures how nonzeros are grouped together and ordered in the target format.
\end{itemize}
As alluded to in \secref{background}, the target tensor format can then be decomposed into level formats that each stores a dimension of the remapped input tensor.
For each of these level formats, the user must then also specify 
\begin{itemize}[leftmargin=1.25\parindent]
  \item what input tensor statistics to compute (in yellow) and 
  \item how to store the coordinates of nonzeros (in blue).
\end{itemize}
To generate code that converts tensors between any two specific formats, our technique combines the aforementioned specifications for the target format with level functions that describe how to iterate over tensors in the source format.
Given the specifications in \figsref{csr-decomposition}{ell-decomposition} as inputs, for instance, our technique generates code like what is shown in \figref{csrell-example}, which performs CSR to ELL conversion.
Just as easily though, given the same specifications in \figref{ell-decomposition} but also level functions that describe how to iterate over COO tensors, our technique instead generates efficient COO to ELL conversion code.
In this way, our technique can generate efficient conversion routines for many combinations of formats without needing specifications for each combination.
%
%

Having a separate language to describe each logical phase provides several benefits.
First, converting to different tensor formats may require similar steps to be taken for only some of the phases, so having each phase be specified separately allows for reuse of the specifications.
For instance, the COO format uses the same data structure as ELL to store column coordinates.
Thus, the two formats can share specifications for assembly (i.e., level functions implemented for the singleton level format) even if they require different coordinate remappings.
Second, having each logical phase be specified separately gives the compiler flexibility to generate code that fuses logically distinct phases only if it is beneficial.
Our technique can thus generate code like~\figref{csrdia-example}, which duplicates and fuses coordinate remapping with the analysis and assembly phases to avoid materializing the offsets of nonzeros.
At the same time, for conversions to formats that store nonzeros in more complex orderings (e.g., Morton order), the compiler can emit code to perform coordinate remapping separately and materialize the additional coordinates.
This eliminates the need to recompute complex remappings.

%% file: sections/coordinate-remapping.tex
As explained in~\secref{tensor-format-conversion}, many efficient sparse tensor conversion algorithms logically transform (i.e., remap) input tensors to higher-order tensors, such that the lexicographic coordinate ordering of nonzeros in the remapped tensors specify how nonzeros are stored in the target formats.
We propose a new language called \emph{coordinate remapping notation}, which precisely describes how a tensor can be remapped so as to capture the various ways that different tensor formats group together and order nonzeros in memory.
We further show how our technique generates code that applies a coordinate remapping to remap the input tensor as part of format conversion.
This eliminates the need for end users to hand-implement additional code that separately performs such a remapping, which the technique of \citeauthor{chou2018}~\shortcite{chou2018} requires for conversions to structured tensor formats.

\subsection{Coordinate Remapping Notation}
\label{sec:tensor-remapping-notation}

\begin{figure}
{\small
\renewcommand{\syntleft}{}
\renewcommand{\syntright}{}
\setlength{\grammarparsep}{0.1em}
\begin{grammar}
<remap_stmt> := <src_indices> `->' <dst_indices> 

<src_indices> := `(' <ivar> \{ `,' <ivar> \} `)'

<dst_indices> := `(' <ivar_let> \{ `,' <ivar_let> \} `)'


<ivar_let> := \{ <var> `=' <ivar_expr> `in' \} <ivar_expr>

<ivar_expr> := <ivar_xor> \{ `|' <ivar_xor> \}

<ivar_xor> := <ivar_and> \{ `^' <ivar_and> \}

<ivar_and> := <ivar_shift> \{ `&' <ivar_shift> \}

<ivar_shift> := <ivar_add> \{ ( `<<' | `>>' ) <ivar_add> \}

<ivar_add> := <ivar_mul> \{ ( `+' | `-' ) <ivar_mul> \}

<ivar_mul> := <ivar_factor> \{ ( `*' | `/' | `\%' ) <ivar_factor> \}

<ivar_factor> := `(' <ivar_expr> `)' | <ivar_counter> | <ivar> | <var> | <const>

<ivar_counter> := `#' \{ <ivar> \}
\end{grammar}
}
  \caption {
    The syntax of coordinate remapping notation.  Expressions in braces may be repeated any number of times.
  }
  \label{fig:remapping-notation-syntax}
\end{figure}

\figref{remapping-notation-syntax} shows the syntax of coordinate remapping notation.
Statements in coordinate remapping notation specify how components in a canonical (non-remapped) input tensor map to components in an output tensor of equal or higher order.
For instance, given a matrix $A$ as input, the statement 
\begin{lstlisting}[xleftmargin=2\parindent]
(i,j) -> (j-i,i,j)
\end{lstlisting}
maps every component $A_{ij}$ to the corresponding component in the $(j-i)$-th slice of three-dimensional remapped tensor.
Applying this remapping to any matrix, which can be stored in any format but must have canonical coordinates $(i,j)$, transforms it to a 3rd-order tensor where each slice contains all nonzeros that lie on the same diagonal in the original matrix.
As \figref{dia-tensorization} shows, the lexicographic coordinate ordering of nonzeros in the resulting tensor precisely reflects the order in which nonzeros are stored in DIA.

Similarly, the BCSR format partitions a matrix into fixed-sized $M \times N$ blocks and stores components of each block contiguously in memory~\cite{bcsr}.
Such grouping of nonzeros can be expressed with the remapping
\begin{lstlisting}[xleftmargin=2\parindent]
(i,j) -> (i/M,j/N,i,j),
\end{lstlisting}
which assigns components that lie within the same block to the same two-dimensional slice (identified by coordinates $(i/M,j/N)$) in the output tensor.

Coordinate remapping notation can express complex tensor reorderings.
The remapping below, for instance, groups together nonzeros that lie within the same constant-sized $N \times N \times N$ block and also orders the blocks as well as the nonzeros within each block in Morton order~\cite{morton1966computer}:
\begin{lstlisting}
(i,j,k) -> 
  (r=i/N in s=j/N in t=k/N in 
     (r&1) | ((s&1)<<1) | ((t&1)<<2) ...,i/N,j/N,k/N,
   u=i%N in v=j%N in w=k%N in 
     (u&1) | ((v&1)<<1) | ((w&1)<<2) ...,i,j,k).
\end{lstlisting}
This remapping exactly describes how the HiCOO tensor format orders nonzeros in memory~\cite{hicoo}.
Nested let expressions are used to first define variables \code{r}, \code{s}, and \code{t} as the coordinates of each block and variables \code{u}, \code{v}, and \code{w} as the coordinates of each nonzero within a block.
The remapping then computes the Morton code of each block and each nonzero within a block by interleaving the bits of those previously defined coordinates using bitwise operations.

\begin{figure}
  \centering
  \includegraphics[width=\linewidth]{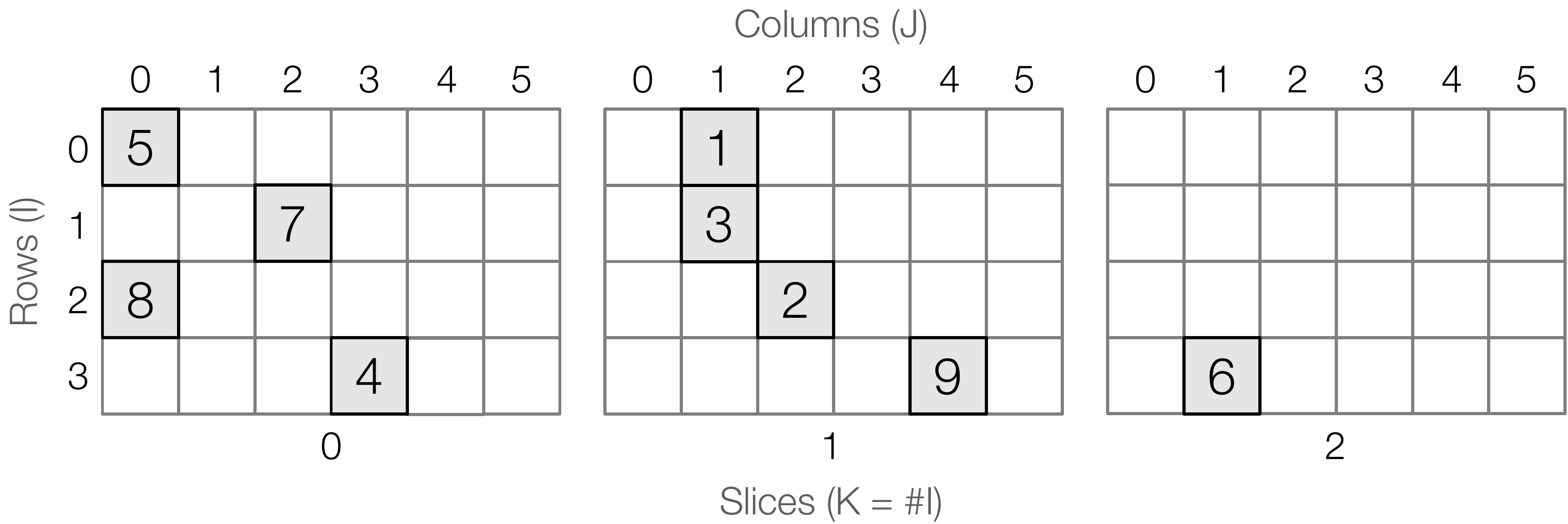}
  \caption {
    Result of applying \code{(i,j) -> (\#i,i,j)} to the matrix in \figref{matrix-example}, assuming nonzeros are iterated over during remapping in the same order as they are stored in \figref{matrix-example-csr}.
  }
  \label{fig:ell-tensorization}
\end{figure}

Coordinate remapping notation also provides counters, denoted by "\code{#}" in~\figref{remapping-notation-syntax}.
Counters  map nonzeros that share the same specified coordinates to distinct slices in the result.
For instance, as \figref{ell-tensorization} shows, the remapping
\begin{lstlisting}[xleftmargin=2\parindent]
(i,j) -> (k=#i in k,i,j)
\end{lstlisting}
assigns the $k$-th nonzero that is iterated over in each row of the input matrix to the $k$-th slice in the output tensor, ensuring nonzeros with the same $i$ coordinate are mapped to distinct slices.
This remapping effectively groups together up to one nonzero from each row of the input matrix, accurately reflecting how formats like ELL and JAD~\cite{Saad1989} store nonzeros.


\subsection{Code Generation}
\label{sec:coordinate-remapping-code-generation}

To support generating sparse tensor conversion routines for arbitrary combinations of formats, we independently annotate each supported format with a coordinate remapping that describes how the format groups together and orders nonzeros.
As the previous examples demonstrate, it is also simple for end users to add support for additional custom formats by using coordinate remapping notation to specify how the new formats lay out nonzeros in memory.
Then, to support converting tensors between two specific formats, our technique emits code that iterates over the input tensor and transforms the (canonical) coordinates of each nonzero by applying the target format's coordinate remapping.


To generate a set of nested loops that efficiently iterate over an input tensor in any source format, our technique uses the method proposed by \citeauthor{kjolstad2017}~\shortcite{kjolstad2017} and generalized by \citeauthor{chou2018}~\shortcite{chou2018}, which is summarized at the end of \secref{background}.
Within the generated loops, our technique then emits code that computes each remapped nonzero's additional coordinates as functions of the nonzero's original coordinates, following the target format's coordinate remapping.

To compute additional coordinates that are defined purely as arithmetic or bitwise expressions of the original coordinates, our technique simply inlines those expressions directly into the emitted code (e.g., lines 6 and 24 in \figref{csrdia-example}, which compute the first coordinate in the output of the remapping \code{(i,j) -> (j-i,i,j)}). 
Remappings that contain let expressions are lowered by first emitting code to initialize the local variables and then inlining the expressions that use those local variables.
For example, a remapped coordinate \code{r=i/N in (r&1) \| ((r&2)<<2)} would be lowered to 
\begin{lstlisting}[xleftmargin=2\parindent,language=c]
int r = i/N;
int m = (r&1) | ((r&2)<<2;
\end{lstlisting}

Coordinate remappings that contain counters are lowered by emitting a counter array for each distinct counter in the remapping.
Each element in the counter array corresponds to a distinct set of coordinates $(i_1,...,i_k)$ that can be used to index into the counter, and the counter array element tracks how many input nonzeros with coordinates $(i_1,...,i_k)$ have been iterated over so far.
Our technique additionally emits code that, for each nonzero having coordinates that correspond to counter array element $c$, first assigns the nonzero to the output tensor slice indexed by $c$ and then increments $c$.
So to apply the remapping \code{(i,j) -> (#i,i,j)} to a COO matrix, for instance, our technique emits the following code:
\begin{lstlisting}[xleftmargin=2\parindent,language=c]
int counter[N] = {0};  // counter array for #i
for (int p = 0; p < nnz; p++) {
  int i = A1_crd[p];
  int j = A2_crd[p];
  int k = counter[i]++;  // k == #i
  // map A(i,j) to coordinates (k,i,j) ...
\end{lstlisting}
If the coordinates used to index into a counter are iterated in order though, our technique reduces the size of the counter array in the generated code by having the counter array be reused across iterations.
This can significantly reduce memory traffic.
For instance, if the input matrix is instead stored in CSR, our technique infers from properties of the format (exposed through the static interface discussed in \secref{background}) that we can efficiently iterate over nonzeros row by row.
Thus, to apply the same coordinate remapping as before, our technique emits optimized code as shown on lines 8--13 in \figref{csrell-example}, which uses the same scalar \code{count} variable to remap the nonzeros in each row of the input matrix.

%% file: sections/attribute-queries.tex
As we also saw in~\secref{tensor-format-conversion}, to avoid having to constantly reallocate and shuffle around stored nonzeros, many efficient tensor conversion algorithms instead allocate memory in one shot based on some statistics about the input tensor.
Computing these statistics, however, requires very different code depending on how the input tensor is stored.
For instance, to convert a matrix to ELL without dynamically resizing the \code{crd} and \code{vals} arrays, one must first determine the maximum number of nonzeros $K$ stored in any row.
If the input matrix is stored in COO, then computing $K$ requires constructing a histogram that records the number of nonzeros in each row, which in turn requires examining all the nonzeros in the matrix.
If the input is stored in CSR, however, then the number of nonzeros in each row can instead be directly computed from the \code{pos} array.
Optimized code for converting CSR matrices to ELL thus does not need to make multiple passes over the input matrix's nonzeros, reducing memory traffic.

We propose a new language called the \emph{attribute query language} that describes statistics of sparse tensors as aggregations over the coordinates of their nonzeros.
The attribute query language is declarative, and attribute queries are specified independently of how the input tensor is actually stored.
This lets our technique lower attribute queries to equivalent sparse tensor computations and then simply leverage prior work on sparse tensor algebra compilation~\cite{kjolstad2017,chou2018,kjolstad2019} to generate optimized code for computing tensor statistics.
As we show in~\secref{tensor-assembly}, our technique can thus generate efficient tensor conversion routines while only requiring users to provide simple-to-specify attribute queries for each potential target format, as opposed to complicated loop nests for every combination of source and target formats.

\subsection{Attribute Query Language}
\label{sec:attribute-query-language}

\begin{figure}
  \centering
  \includegraphics[width=\linewidth]{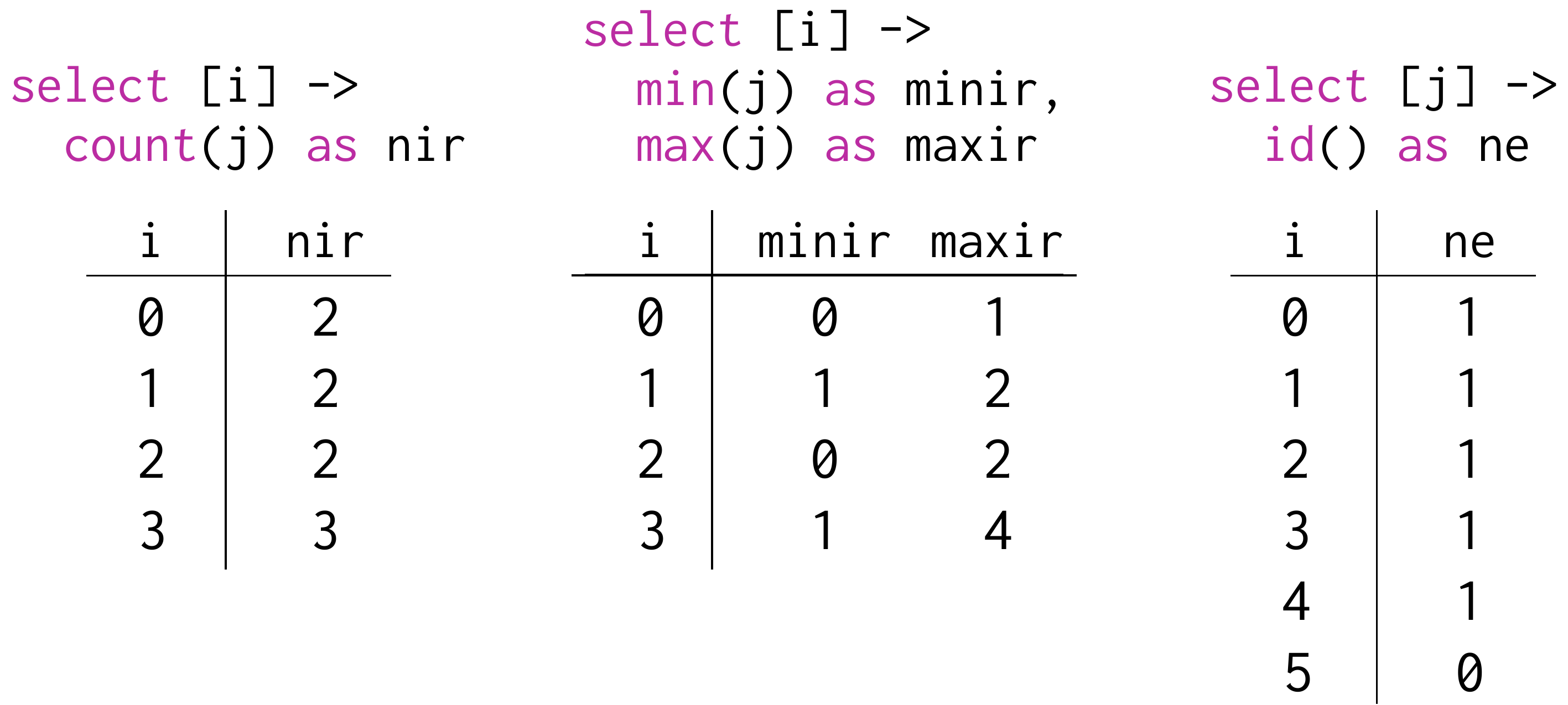}
  \caption{
    Examples of attribute queries computed on the tensor shown in \figref{matrix-example}.
  }
  \label{fig:attribute-query-examples}
\end{figure}

The attribute query language lets users compute summaries of a tensor's sparsity structure by performing aggregations over the coordinates of the tensor's nonzeros.
All queries in the attribute query language take the form
\begin{lstlisting}[xleftmargin=2\parindent,language=sql,mathescape,morekeywords={id}]
select [i$_1$,...,i$_m$] -> 
  <aggr$_1$> as label$_1$, ..., <aggr$_n$> as label$_n$
\end{lstlisting}
where each \code{i}$_k$ denotes a coordinate into dimension $I_k$ of some $r$-dimensional tensor $A$ and \code{<aggr}$_k$\code{>} invokes the aggregation function \code{count}, \code{max}, \code{min}, or \code{id}.
The result of an attribute query is conceptually a map that, for every distinct set of coordinates $(i_1, \ldots, i_m)$, stores computed statistics about the $I_{m+1} \times \cdots \times I_r$ subtensor $A'$ in $A$ identified by those coordinates.
\figref{attribute-query-examples} shows examples of different attribute queries computed on the same tensor.

\code{count(}i$_{m+1}$\code{,...,i}$_l$\code{)} computes the number of nonzero $I_{l+1} \times \cdots \times I_r$ subtensors, each of which can be identified by a distinct set of coordinates $(i_1, \ldots, i_l)$, that are contained in $A'$.
For instance, if $I$, $J$, $K$ represent the slice, row, and column dimensions of a 3rd-order tensor $B$, then the query 
\begin{lstlisting}[xleftmargin=2\parindent,language=sql,mathescape,morekeywords={id}]
select [i] -> count(j) as nnr_in_slice
\end{lstlisting}
computes the number of nonzero rows contained in each $J \times K$ slice of $B$, while the query 
\begin{lstlisting}[xleftmargin=2\parindent,language=sql,mathescape,morekeywords={id}]
select [i] -> count(j,k) as nnz_in_slice
\end{lstlisting}
computes the number of nonzeros in each $J \times K$ slice.
\figref{attribute-query-examples} (left) shows how \code{count} queries can be used to compute the number of nonzeros in each row of a matrix, which is required when converting it from COO to CSR for instance.

\code{max(i}$_{m+1}$\code{)} and \code{min(i}$_{m+1}$\code{)} compute, for each subtensor $A'$, the largest and smallest coordinates $i_{m+1}$ such that the $i_{m+1}$-th slice of $A'$ along dimension $I_{m+1}$ is nonzero.
For instance, if \code{Q} is the result of the query in \figref{attribute-query-examples} (middle), then \code{Q[3].minir == 1} and \code{Q[3].maxir == 4} since all nonzeros in row 3 of the tensor in \figref{matrix-example} lie between columns 1 and 4.

Finally, \code{id} simply returns 1 if a subtensor $A'$ contains nonzeros and 0 otherwise.
So if \code{R} is the result of the query in \figref{attribute-query-examples} (right), then \code{R[4].ne == 1} since column 4 contains a nonzero while \code{R[5].ne == 0} since the last column is empty.

The attribute query language can be used with coordinate remapping notation to compute even more complex attributes of structured tensors.
For example, let $A$ be a $K \times I \times J$ tensor obtained by applying the remapping \code{(i,j) ->} \code{(j-i,i,j)} to a matrix $B$.
Since each slice of $A$ along dimension $K$ corresponds to a unique diagonal in $B$, computing 
\begin{lstlisting}[xleftmargin=2\parindent,language=sql,mathescape,morekeywords={id}]
select [k] -> id() as ne
\end{lstlisting}
on $A$ results in a bit set that encodes the set of all nonzero diagonals in $B$.
This, as mentioned in \secref{tensor-format-conversion}, is precisely information that would be required if one were to convert $B$ from CSR to DIA.
Furthermore, since the coordinate of each slice of $A$ is defined to be the offset of the corresponding diagonal in $B$ from the main diagonal, applying the query 
\begin{lstlisting}[xleftmargin=2\parindent,language=sql,mathescape,morekeywords={id}]
select [] -> min(k) as lb, max(k) as ub
\end{lstlisting}
to $A$ computes the lower and upper bandwidths of matrix $B$.

\subsection{Code Generation}
\label{sec:attribute-query-language-code-generation}

\begin{table*}
  \caption{
    Example transformations that our technique applies to optimize attribute queries.
    We augment level formats with a property that specifies if a dimension stores explicit zeros, which lets our technique determine if a tensor stores only nonzeros.
  }
  \centering
  {\footnotesize
  \begin{tabularx}{\textwidth}{lXp{3in}}
    \toprule
    \multicolumn{1}{c}{Transformation} & \multicolumn{1}{c}{Definition} & \multicolumn{1}{c}{Preconditions and Postconditions} \\
    \midrule
    reduction-to-assign &
    {$\!\begin{aligned}[t]
      & \big( \iteration{j_1} \cdots \iteration{j_n} A_{i_1 \cdots i_m} \reduce{\oplus} expr \big) \\[-0.2em]
      & \implies \big( \iteration{j_1} \cdots \iteration{j_n} A_{i_1 \cdots i_m} \assign expr \big)
    \end{aligned}$} &
    For each $j_k$, there exists an $i_l$ such that $j_k \equiv i_l$.
    $\oplus$ is any reduction operator. 
    $A$ is initialized to the zero tensor. \\
    \midrule
    inline-temporary &
    {$\!\begin{aligned}[t]
      & \big(\iteration{i_1} \cdots \iteration{i_m} A_{i_1 \cdots i_l} \reduce{\oplus} f(W_{i_1 \cdots i_m}) \big) \\[-0.2em]
      &  \where \big(\iteration{j_1} \cdots \iteration{j_n} W_{i_1 \cdots i_m} \assign expr \big) \\[-0.2em]
      & \implies \big(\iteration{j_1} \cdots \iteration{j_n} A_{i_1 \cdots i_l} \reduce{\oplus} f(expr) \big)
    \end{aligned}$} &
    $f$ is any function that takes only $W$ as tensor operand. $\oplus$ is any reduction operator or a simple assignment.
    \\
    \midrule
    simplify-width-count & 
    {$\!\begin{aligned}[t]
      & \big( \iteration{j_1} \cdots \iteration{j_n} A_{i_1 \cdots i_m} \reduce{+} \map(B_{j_1 \cdots j_n}, c) \big) \\[-0.2em]
      & \implies \big( \iteration{j_1} \cdots \iteration{j_{n-1}} A_{i_1 \cdots i_m} \reduce{+} B'_{j_1 \cdots j_{n-1}} \cdot c \big)
    \end{aligned}$} &
    $B$ stores only nonzeros, and $j_n$ is a reduction variable that indexes into the innermost dimension of $B$ (i.e., $J_n$). 
    $c$ is any constant.
    $B'$ is a tensor that encodes the number of nonzeros in each slice of $B$ indexed by coordinates $(j_1, ..., j_{n-1})$; values of $B'$ are not materialized but dynamically computed with calls to level functions that define iteration over dimension $J_n$ of $B$.
    \\
    \midrule
    counter-to-histogram &
    {$\!\begin{aligned}[t]
      & \big( \iteration{j_1} \cdots \iteration{j_n} A_{i_1 \cdots i_m} \reduce{\max} \map(B_{j_1 \cdots j_n}, \#j_k \cdots j_l + 1) \big) \\[-0.2em]
      & \implies \big(\iteration{i_1} \cdots \iteration{i_l} A_{i_1 \cdots i_m} \reduce{\max} W_{i_1 \cdots i_l} \big) \\[-0.2em]
      &  \where \big(\iteration{j_1} \cdots \iteration{j_n} W_{i_1 \cdots i_m j_k \cdots j_l} \reduce{+} \map(B_{j_1 \cdots j_n}, 1) \big)
    \end{aligned}$} &
    None.
    \\
    \bottomrule
  \end{tabularx}
  }
  \label{tab:attribute-query-optimizations}
\end{table*}

\citeauthor{kjolstad2019}~\shortcite{kjolstad2019} introduce concrete index notation, which is a language for precisely specifying tensor computations.
For instance, an operation that computes the sum of every row in a matrix $A$ can be expressed as $\titeration{i}\iteration{j} x_i \reduce{+} A_{ij}$, where each $\forall$ specifies iteration over a dimension of $A$.
\citeauthor{kjolstad2019} show how computations expressed in concrete index notation can be lowered to efficient imperative code; we refer readers to Section 5 in~\cite{kjolstad-thesis} for more details.
At a high level, this is done recursively dimension by dimension in the order specified by the $\forall$s.
To generate code for the previous example, for instance, a compiler would first emit a loop to iterate over all rows of $A$.
Within that loop, the compiler would then emit a second loop to iterate over all columns of $A$ in order to compute the sum of nonzeros in row $i$.
Again, specializing the emitted code to operands in arbitrary formats can be done in the same way described at the end of \secref{background}.

To generate efficient code that computes an attribute query, our technique simply reformulates the query as sparse tensor algebra computation.
The query is first lowered to a canonical form in concrete index notation, which we extend with the ability to index into results using coordinates that are computed as arbitrary functions of index variables.
The canonical form of the query is subsequently optimized by applying a set of predefined transformations to simplify the computation.
Finally, the optimized query in concrete index notation is compiled to imperative code by straightforwardly leveraging the techniques of \citeauthor{kjolstad2017} and \citeauthor{chou2018} as summarized above.
This approach works as long as query results are stored in a format, like dense arrays, that can itself be efficiently assembled without needing attribute queries.

More precisely, let $A$ be an $I_1 \times \cdots \times I_r$ tensor obtained by applying some remapping to a $J_1 \times \cdots \times J_n$ tensor $B$.
Then, to compute an attribute query of the form 
\begin{lstlisting}[xleftmargin=2\parindent,language=sql,mathescape,morekeywords={id}]
select [i$_1$,...,i$_m$] -> id() as Q
\end{lstlisting}
on $A$ for instance, our technique lowers the query to its canonical form in concrete index notation as 
$$
\titeration{j_1} \cdots \iteration{j_n} Q_{i_1 \cdots i_m} \reduce{|} \map(B_{j_1 \cdots j_n}, 1),
$$
where $\reduce{|}$ denotes Boolean OR reduction.
The computation above logically iterates over every component of $B$, computes the coordinates $(i_1, \ldots, i_m)$ of each component $B_{j_1 \cdots j_n}$ in the remapped tensor $A$, and sets the corresponding component in the Boolean result tensor $Q$ to true (1).
(All components of $Q$ are assumed to be initialized to false.)
The map operator returns the second argument if the first argument is nonzero (or true) and zero otherwise, which ensures only the coordinates of nonzeros in $B$ are aggregated.
So if, for instance, $C$ is a $K \times I \times J$ tensor obtained by applying the remapping \code{(i,j) -> (j-i,i,j)} to a matrix $D$, then to compute \query{select [k] -> id() as Q} on $C$, our technique lowers the query to the computation $\titeration{i}\iteration{j} Q_{j-i} \reduce{|} \map(D_{ij}, 1).$
For each nonzero of $D$, this computation computes the nonzero's offset from the main diagonal and sets the corresponding component in $Q$ to true.
The query result $Q$ thus strictly encodes the set of diagonals in $D$ that contain nonzeros.

In a similar way, our technique lowers \code{count} queries 
\begin{lstlisting}[xleftmargin=2\parindent,language=sql,mathescape,morekeywords={id}]
select [i$_1$,...,i$_m$] -> count(i$_{m+1}$,...,i$_l$) as Q
\end{lstlisting}
on $A$ to their canonical form 
\begin{align*}
&\big(\titeration{i_1} \cdots \iteration{i_l} Q_{i_1 \cdots i_m} \reduce{+} \map(W_{i_1 \cdots i_l}, 1) \big) \\[-0.2em]
&\where \big(\titeration{j_1} \cdots \iteration{j_n} W_{i_1 \cdots i_l} \reduce{|} \map(B_{j_1 \cdots j_n}, 1) \big).
\end{align*}
The computation above first iterates over the nonzeros of $B$ to compute the intermediate result $W$, which encodes whether each subtensor of $A$ identified by coordinates $(i_1, \ldots, i_l)$ is nonzero.
The computation then sums over dimensions $I_{m+1}$ through $I_l$ of $W$ to compute the number of aforementioned subtensors that are nonzero and contained in each higher-order subtensor with coordinates $(i_1, \ldots, i_m)$.

Our technique also generates code for \code{max} queries 
\begin{lstlisting}[xleftmargin=2\parindent,language=sql,mathescape,morekeywords={id}]
select [i$_1$,...,i$_m$] -> max(i$_{m+1}$) as Q
\end{lstlisting}
by lowering them to their canonical form 
$$
\titeration{j_1} \cdots \iteration{j_n} Q'_{i_1 \cdots i_m} \reduce{\max} \map(B_{j_1 \cdots j_n}, i_{m+1} - s + 1),
$$
where $s$ denotes the smallest possible coordinate along dimension $I_{m+1}$.
$Q'$ is assumed to be initialized to the zero tensor, so by mapping each input tensor component to its remapped coordinate $i_{m+1}$ plus the constant $(1-s)$, we ensure that only the coordinates of nonzeros are actually aggregated.
$Q'$ can thus be interpreted as the actual result of the original query (i.e., $Q$) but just shifted by $(1-s)$; in other words, $Q_{i_1 \cdots i_m} \equiv Q'_{i_1 \cdots i_m} + s - 1$.
Similarly, \code{min} queries 
\begin{lstlisting}[xleftmargin=2\parindent,language=sql,mathescape,morekeywords={id}]
select [i$_1$,...,i$_m$] -> min(i$_{m+1}$) as Q
\end{lstlisting}
are lowered to their canonical form 
$$
\titeration{j_1} \cdots \iteration{j_n} Q'_{i_1 \cdots i_m} \reduce{\max} \map(B_{j_1 \cdots j_n}, {-i_{m+1}} + t + 1),
$$
where $t$ denotes the largest possible coordinate along dimension $I_{m+1}$ and $Q'$ is the query result but negated and shifted by $(t + 1)$; in other words, $Q_{i_1 \cdots i_m} \equiv -Q'_{i_1 \cdots i_m} + t + 1$.

After an attribute query is lowered to its canonical form, our technique eagerly applies a set of predefined transformations on the query computation to optimize its performance.
\tabref{attribute-query-optimizations} shows a subset of transformations that our technique uses.
In general, these transformations exploit properties of the input tensor and its underlying storage format to reduce the number of dimensions that have to be iterated and to eliminate redundant temporaries. 

To see how our technique optimizes attribute queries, consider the example query \query{select [i] -> count(j) as Q} applied to an $I \times J$ matrix $B$.
As described before, our technique first lowers this query to its canonical form
$$
\big(\titeration{i} \iteration{j} Q_i \reduce{+} \map(W_{ij}, 1) \big) \where \big(\titeration{i} \iteration{j} W_{ij} \reduce{|} \map(B_{ij}, 1) \big).
$$
Our technique then proceeds to iteratively and eagerly apply the transformations shown in~\tabref{attribute-query-optimizations} on the computation above.
In particular, each iteration variable bound to a $\forall$ is used to independently index into a dimension of $W$, so the substatement that defines $W$ satisfies the preconditions of the reduction-to-assign transformation.
Our technique thus applies the aforementioned transformation on the substatement that defines $W$ to obtain  
$$
\big(\titeration{i} \iteration{j} Q_i \reduce{+} \map(W_{ij}, 1) \big) \where \big(\titeration{i} \iteration{j} W_{ij} \assign \map(B_{ij}, 1) \big).
$$
Then, since the temporary $W$ is no longer the result of a reduction operation, our technique eliminates it by applying the inline-temporary transformation to obtain  
$$
\titeration{i} \iteration{j} Q_{i} \reduce{+} \map(\map(B_{ij}, 1), 1),
$$
which is then trivially rewritten to $\titeration{i} \iteration{j} Q_{i} \reduce{+} \map(B_{ij}, 1)$ by applying constant folding.
If $B$ is stored in COO, then we can directly apply the techniques of \citeauthor{kjolstad2017} and \citeauthor{chou2018} to lower this rewritten statement down to imperative code shown on lines 1--6 in \figref{coocsr-example}.
However, if $B$ is stored in CSR (with only nonzeros stored), then our technique additionally applies the simplify-width-count transformation followed by reduction-to-assign again to get the final query 
$$
\iteration{i} Q_{i} = B'_{i},
$$
where each component of $B'$ is dynamically computed as \code{pos[i+1] - pos[i]}.
The optimized query thus avoids iterating over $B$'s nonzeros, thereby reducing memory traffic.


%% file: sections/assembly-abstraction.tex
\begin{figure*}
  \centering
  \includegraphics[width=0.95\linewidth]{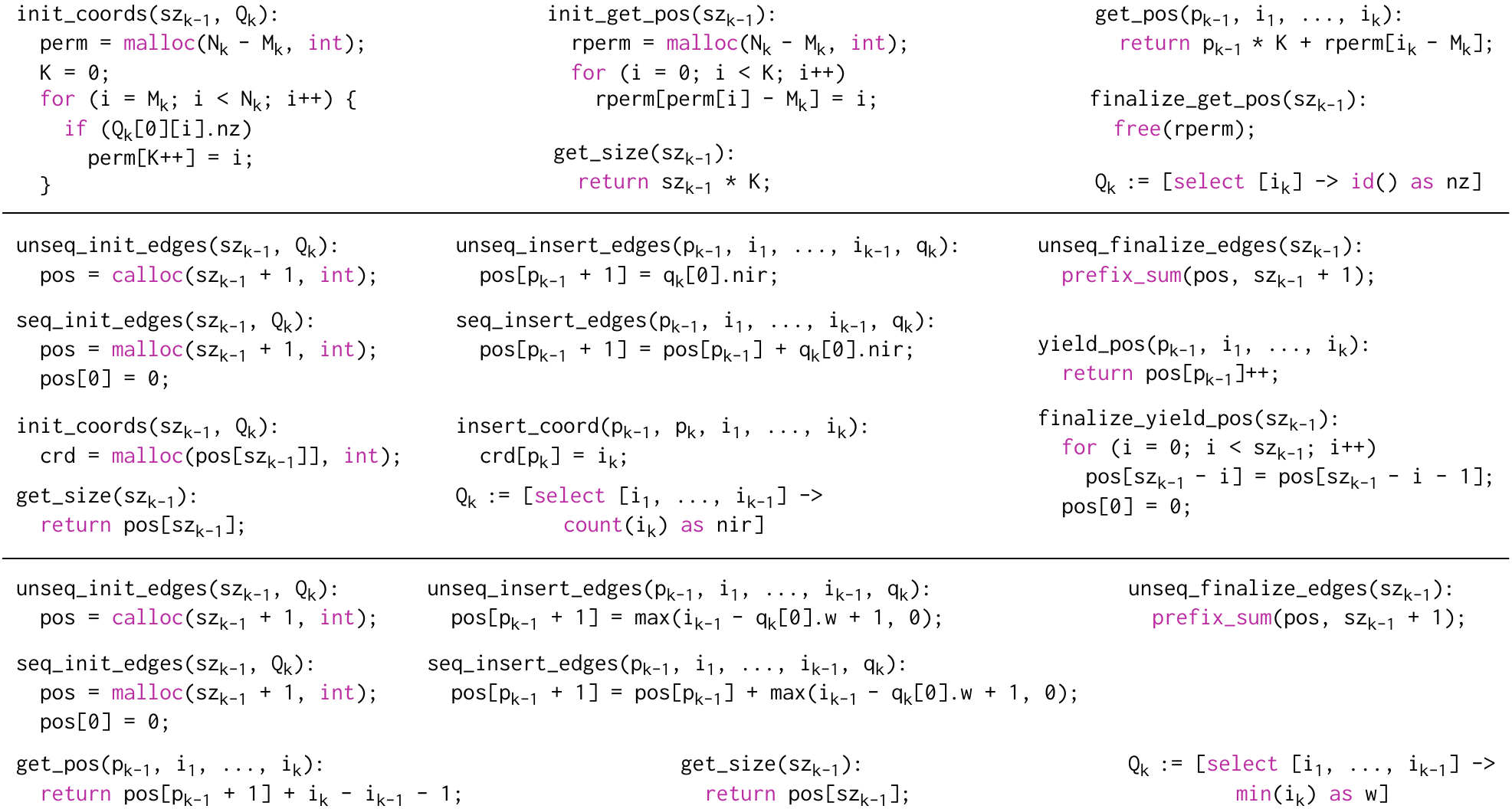}
  \caption{
    Implementations of the assembly abstraction, including level function definitions and the required attribute queries, for three different level formats: \emph{squeezed} (top), \emph{compressed} (middle), and \emph{banded} (bottom).
    \emph{Squeezed} stores the dimension of offsets in (remapped) DIA tensors.
    \emph{Compressed} stores the column dimension of CSR tensors as well as the row dimension of COO tensors.
    \emph{Banded} stores the column dimension of tensors in the skyline format~\cite{mkl}, which stores all components between the first nonzero and the diagonal for every row.
    $M_k$ and $N_k$ denote the lower and upper bounds of the $k$-th dimension.
  }
  \label{fig:assembly-function-examples}
\end{figure*}

As explained in~\secref{background}, a sparse tensor can be modeled as a hierarchical structure of coordinates, where each stored component is represented by a path from the root to a leaf.
We can thus view any tensor format simply as a composition of level formats that each stores a level of a coordinate hierarchy.
This abstraction lets us reason about sparse tensor assembly as coordinate hierarchy construction.

We extend the coordinate hierarchy abstraction with new primitives (level functions) that describe how each level can be efficiently constructed (assembled).
These new level functions, unlike analogous ones proposed by~\citeauthor{chou2018}~\cite{chou2018}, describe how coordinates and edges can be efficiently inserted into a coordinate hierarchy assuming certain statistics about the input tensor have been precomputed.

\figsref{ell-decomposition}{assembly-function-examples} show how level formats that use disparate data structures can implement the new assembly level functions.
All these implementations expose the same static interface, which lets our code generator reason about and emit efficient code to convert tensors between a wide range of formats.
And by using the same level formats to express different tensor formats that share common data structures, our technique can reuse the same level function implementations to generate conversion routines for many different target formats.
For example, the column dimensions of CSR tensors and the row dimensions of COO tensors can both be stored using the same (i.e., compressed) level format.
The code generator can thus use the same implementations of the new assembly level functions (for the compressed level format) to generate parts of code that convert tensors to either CSR or COO.
This limits the one-time effort needed to implement our extended coordinate hierarchy abstraction.


\subsection{Assembly Abstraction}
\label{sec:assembly-abstraction}

Our extended coordinate hierarchy abstraction assumes that coordinate hierarchies can be constructed level by level from top to bottom.
The assembly of each level is decomposed into two logical phases: \emph{edge insertion} and \emph{coordinate insertion}.

The edge insertion phase, which is optional, logically bulk inserts edges into a coordinate hierarchy to connect coordinates in adjacent levels.
Edge insertion models the assembly of data structures that map nonzeros to their containing subtensors.
Depending on whether each position (node) in the preceding parent level can be iterated in sequence, edge insertion can be done in a \emph{sequenced} or \emph{unsequenced} fashion.

Unsequenced edge insertion is defined in terms of three level functions that any level format may implement:
\begin{itemize}[leftmargin=1.25\parindent]
  \item \code{unseq_init_edges(sz}$_{k-1}$\code{, Q}$_k$\code{) -> void}
  \item \code{unseq_insert_edges(p}$_{k-1}$\code{, i}$_1$\code{, ..., i}$_{k-1}$\code{, q}$_k$\code{) -> void}
  \item \code{unseq_finalize_edges(sz}$_{k-1}$\code{) -> void}
\end{itemize}
$Q_k$ denotes the (complete) results of attribute queries that a level format specifies must be precomputed, while $q_k$ denotes only the elements of $Q_k$ indexed by coordinates $(i_1, \ldots, i_{k-1})$.
$sz_{k-1}$ is the size of the parent level and can be computed as a function of its own parent's size by calling the level function 
\begin{lstlisting}[xleftmargin=1.25\parindent,language=c,mathescape]
get_size(sz$_{k-1}$) -> sz$_k$,
\end{lstlisting}
which all level formats must also implement.
\code{unseq_init_} \code{edges} initializes data structures that the level format uses to logically store edges.
Then, for each position $p_{k-1}$ (which represents a subtensor with coordinates $(i_1, \ldots, i_{k-1})$) in the parent level, \code{unseq_insert_edges} allocates some number of child coordinates to be connected to $p_{k-1}$.
The number of child coordinates allocated can be computed as any arbitrary function of $q_k$.
Finally, \code{unseq_finalize_edges} inserts edges into the coordinate hierarchy so that each coordinate in the parent level is connected to as many children as it was previously allocated.
\figref{assembly-pseudocode} (left) shows how these level functions can logically be invoked to bulk insert edges.

\begin{figure*}
  \centering
  \begin{minipage}[t]{0.49\linewidth}
    \centering
    \begin{tabular}{c}
      \begin{lstlisting}[xleftmargin=2\parindent,language=c,mathescape]
sz$_{k-1}$ = get_size$_{k-1}$(get_size$_{k-2}$($\cdots$(1)$\cdots$));
unseq_init_edges(sz$_{k-1}$, Q$_k$);
for (position $p_{k-1}$ in parent level | $\exists$ coords 
      i$_1$, ..., i$_{k-1}$ connecting $p_{k-1}$ to root) {
  q$_k$[:] = Q$_k$[:][i$_1$, ..., i$_{k-1}$];
  unseq_insert_edges(p$_{k-1}$, i$_1$, ..., i$_{k-1}$, q$_k$);
}
unseq_finalize_edges(sz$_{k-1}$);
      \end{lstlisting}
    \end{tabular}
  \end{minipage}
  \hfill
  \begin{minipage}[t]{0.49\linewidth}
    \centering
    \begin{tabular}{c}
      \begin{lstlisting}[xleftmargin=2\parindent,language=c,mathescape]
init_coords(sz$_{k-1}$, Q$_k$);
init_{get|yield}_pos(sz$_{k-1}$);
for (nonzero with coords i$_1$, ..., i$_k$) {
  for (j = 1; j <= k; j++)  // can be unrolled
    p$_j$ = {get|yield}_pos$_j$(p$_{j-1}$, i$_1$, ..., i$_j$);
  insert_coord(p$_{k-1}$, p$_k$, i$_1$, ..., i$_k$);
}
finalize_{get|yield}_pos(sz$_{k-1}$);
      \end{lstlisting}
    \end{tabular}
  \end{minipage}
  \caption {
    Unsequenced edge insertion (left) and coordinate insertion (right), expressed in terms of calls to level functions.
  }
  \label{fig:assembly-pseudocode}
\end{figure*}

Sequenced edge insertion, by contrast, is defined in terms of just two level functions:
\begin{itemize}[leftmargin=1.25\parindent]
  \item \code{seq_init_edges(sz}$_{k-1}$\code{, Q}$_k$\code{) -> void}
  \item \code{seq_insert_edges(p}$_{k-1}$\code{, i}$_1$\code{, ..., i}$_{k-1}$\code{, q}$_k$\code{) -> void}

\end{itemize}
These level functions are analogous to \code{unseq_init_edges} and \code{unseq_insert_edges} and can be invoked in similar ways.
Sequenced edge insertion, however, assumes that all positions in the parent level are iterated in order.
Thus, \code{seq_insert_} \code{edges} is responsible for both allocating the appropriate number of child coordinates to each parent and actually inserting the edges, and a separate \code{finalize} function is not necessary.

The coordinate insertion phase logically iterates over the input tensor's nonzeros and inserts their coordinates into a coordinate hierarchy.
This phase models the assembly of data structures that store the coordinates and values of the nonzeros, and it is defined in terms of five level functions:
\begin{itemize}[leftmargin=1.25\parindent]
  \item \code{init_coords(sz}$_{k-1}$\code{, Q}$_k$\code{) -> void}
  \item \code{init_\{get\|yield\}_pos(sz}$_{k-1}$\code{) -> void}
  \item \code{\{get\|yield\}_pos(p}$_{k-1}$\code{, i}$_1$\code{, ..., i}$_k$\code{) -> p}$_k$
  \item \code{insert_coord(p}$_{k-1}$\code{, p}$_k$\code{, i}$_1$\code{, ..., i}$_k$\code{) -> void}
  \item \code{finalize_\{get\|yield\}_pos(sz}$_{k-1}$\code{) -> void}
\end{itemize}
\code{init_coords} allocates and initializes data structures for storing coordinates in a coordinate hierarchy level.
If a level format implicitly encodes coordinates (e.g., as a fixed range) using some fixed set of parameters, then \code{init_coords} also compute those parameters as functions of the attribute query results $Q_k$.
On the other hand, if a level format explicitly stores coordinates in memory, then the coordinates of nonzeros are inserted by invoking \code{insert_coord} for each nonzero.
The position $p_k$ at which each nonzero should be inserted is computed by invoking either \code{get_pos} or \code{yield_pos}.
The former guarantees that nonzeros with the same coordinates are inserted at the same position.
The latter allows duplicate coordinates to be inserted at different positions.
Both functions, however, may rely on auxiliary data structures to track where to insert coordinates; \code{init_\{get\|yield\}_pos} and \code{finalize_\{get\|yield\}_pos} initializes and cleans up those data structures.
\figref{assembly-pseudocode} (right) shows how all these level functions can be invoked to perform coordinate insertion.

\subsection{Code Generation}
\label{sec:assembly-code-generation}

To generate code that converts sparse tensors between two formats, our code generator emits loops that iterate over a tensor in the source format and apply the target format's coordinate remapping to each nonzero.
This is done by applying the technique described in~\secref{coordinate-remapping-code-generation}.
Then, within each loop nest that iterates over the (remapped) input tensor, the code generator emits code that invokes the level functions described in~\secref{assembly-abstraction} to store each nonzero into the result.
The emitted code is finally specialized to the target format by inlining its implementations of the aforementioned level functions (e.g., as shown in \figref{assembly-function-examples}).
This approach enables the code generator to support disparate target (and source) formats.
At the same time, it limits the complexity of the code generator, since the code generator does not need to hard-code for specific data structures but can simply reason about how to invoke a fixed set of level functions.

In order to minimize memory traffic at runtime, our technique emits code that, wherever possible, fuses the assembly of adjacent levels in the result coordinate hierarchy.
Adjacent levels can be assembled together as long as only the parent level requires a separate edge insertion phase (or if none do).
As an example, none of the level formats that compose DIA requires edge insertion.
Thus, our technique will emit code to convert any matrix to DIA by iterating over the matrix just once and assembling all output dimensions (levels) together.

For each set of levels that can be assembled together, our technique then simply has to emit code like what is shown in \figref{assembly-pseudocode} to perform edge insertion (if required) followed by coordinate insertion.
If a level format implements both variants of edge insertion, then our technique selects one based on whether the parent level can be iterated in order.
The code generator infers this from properties exposed through the coordinate hierarchy abstraction that specify if the parent level stores coordinates in order.
If a level format implements \code{yield_pos} but does not permit storing duplicates of the same coordinate, our technique also emits logic to perform deduplication on the fly by keeping track of inserted coordinates.

To see how our technique works, suppose we are generating code to convert COO tensors to CSR. 
To obtain code that assembles the column dimension of the result, the code generator first emits sequenced edge insertion code that has the same structure as \figref{assembly-pseudocode} (left), except with all level functions replaced by their sequenced counterparts.
The emitted code is then specialized to CSR by replacing the level function calls with the compressed level format's implementations of those functions (\figref{assembly-function-examples}, middle).
The result is lines 7--11 in \figref{coocsr-example}, which iterate over all rows of the output in order and reserve enough memory to store each row's nonzeros.
In a similar way, the code generator emits code in \figref{assembly-pseudocode} (right) to perform coordinate insertion and then specializes it to CSR, yielding lines 12--25 in \figref{coocsr-example}.



%% file: sections/evaluation.tex
{
\setlength{\tabcolsep}{4.9pt}
\begin{table}
  \caption{
    Statistics about matrices used in our experiments.
    Non-symmetric matrices are highlighted in gray.
  }
  \centering
  {\footnotesize
  \begin{tabular}{llrrr}
    \toprule
    \multicolumn{1}{c}{Matrix} & \multicolumn{1}{c}{Dimensions} & \multicolumn{1}{c}{NNZ} & \multicolumn{1}{c}{\begin{tabular}{@{}c@{}}Nonzero \\ Diagonals \end{tabular}} & \multicolumn{1}{c}{\begin{tabular}{@{}c@{}}Max. \\ NNZ/row\end{tabular}} \\
    \midrule
    pdb1HYS & 36.4K $\times$ 36.4K & 4.34M & 26K & $\num{204}$ \\
    jnlbrng1 & 40.0K $\times$ 40.0K & 199K & $\num{5}$ & $\num{5}$ \\
    obstclae & 40.0K $\times$ 40.0K & 199K & $\num{5}$ & $\num{5}$ \\
    \rowcolor{nonsymmcolor}
    chem\_master1 & 40.4K $\times$ 40.4K & 201K & $\num{5}$ & $\num{5}$ \\
    \rowcolor{nonsymmcolor}
    rma10 & 46.8K $\times$ 46.8K & 2.37M & 17K & $\num{145}$ \\
    dixmaanl & 60.0K $\times$ 60.0K & 300K & $\num{7}$ & $\num{5}$ \\
    cant & 62.5K $\times$ 62.5K & 4.01M & $\num{99}$ & $\num{78}$ \\
    \rowcolor{nonsymmcolor}
    shyy161 & 76.5K $\times$ 76.5K & 330K & $\num{7}$ & $\num{6}$ \\
    consph & 83.3K $\times$ 83.3K & 6.01M & 13K & $\num{81}$ \\
    denormal & 89.4K $\times$ 89.4K & 1.16M & $\num{13}$ & $\num{13}$ \\
    \rowcolor{nonsymmcolor}
    Baumann & 112K $\times$ 112K & 748K & $\num{7}$ & $\num{7}$ \\
    cop20k\_A & 121K $\times$ 121K & 2.62M & 221K & $\num{81}$ \\
    shipsec1 & 141K $\times$ 141K & 3.57M & 10K & $\num{102}$ \\
    \rowcolor{nonsymmcolor}
    majorbasis & 160K $\times$ 160K & 1.75M & $\num[group-separator={,}]{22}$ & $\num{11}$  \\
    \rowcolor{nonsymmcolor}
    scircuit & 171K $\times$ 171K & 959K & 159K & $\num{353}$ \\
    \rowcolor{nonsymmcolor}
    mac\_econ\_fwd500 & 207K $\times$ 207K & 1.27M & $\num{511}$ & $\num{44}$ \\
    pwtk & 218K $\times$ 218K & 11.5M & 20K & $\num{180}$ \\
    Lin & 256K $\times$ 256K & 1.77M & $\num{7}$ & $\num{7}$ \\
    ecology1 & 1.00M $\times$ 1.00M & 5.00M & $\num{5}$ & $\num{5}$ \\
    \rowcolor{nonsymmcolor}
    webbase-1M & 1.00M $\times$ 1.00M & 3.11M & 564K & $\num{4700}$ \\
    \rowcolor{nonsymmcolor}
    atmosmodd & 1.27M $\times$ 1.27M & 8.81M & $\num{7}$ & $\num{7}$ \\
    \bottomrule
  \end{tabular}
  }
  \label{tab:input-summary}
\end{table}
}

We evaluate our technique and find that it generates efficient sparse tensor conversion routines for many combinations of disparate source and target formats.
The generated conversion routines have performance similar to or better than equivalent hand-implemented versions.
We also find that, for combinations of source and target formats that are not directly supported by a library, our technique can further optimize conversions between those formats by emitting code that avoids materializing temporaries.


\subsection{Experimental Setup}
\label{sec:experimental-setup}

We implemented a prototype of our technique as extensions to the open-source taco tensor algebra compiler~\cite{kjolstad2017}.
To evaluate it, we compare code that our technique generates against hand-implemented versions in SPARSKIT~\cite{Saad94sparskit}, a widely used~\cite{PERATTA200642,jiang2007} Fortran sparse linear algebra library, and Intel MKL~\cite{mkl}, a C and Fortran math processing library that is optimized for Intel processors.
We also evaluate our technique against taco without our extensions.\footnote{\url{https://github.com/tensor-compiler/taco/tree/c0e93b65}}

All experiments are conducted on a 2.5 GHz Intel Xeon E5-2680 v3 machine with 30 MB of L3 cache and 128 GB of main memory.
The machine runs Ubuntu 18.04.3 LTS.
We compile code that our technique generates using GCC 7.5.0 and build SPARSKIT from source using GFortran 7.5.0.
We run each experiment 50 times under cold cache conditions and report median serial execution times.

We run our experiments with real-world matrices of varying sizes and structures as inputs.
These matrices, listed in \tabref{input-summary}, come from applications in disparate domains and are obtained from the SuiteSparse Matrix Collection~\cite{suitesparse}.

\subsection{Performance Evaluation}
\label{sec:performance-evaluation}

{
\setlength{\tabcolsep}{1.95pt}
\setlength{\cmidrulekern}{0.2em}
\newcolumntype{o}{>{\columncolor{nonsymmcolor}}r}
\begin{table*}
  \caption{
    Normalized execution times of conversion routines that are implemented or generated in SPARSKIT (skit), Intel MKL (mkl) and taco without our extensions (taco w/o ext.), relative to code that our technique generates (taco w/ ext.).
    The actual execution times (in milliseconds) of code that our technique generates are shown in parentheses.
    For CSR to CSC conversion, we only show results for nonsymmetric matrices (rows highlighted in gray) since CSR and CSC are equivalent for symmetric matrices.
    For symmetric matrices, we cast CSC to DIA/ELL conversion as CSR to DIA/ELL conversion and report the same results.
    For conversions to DIA/ELL, we also omit results for matrices that would have to be stored with more than 75\% of values being zeros, since having to compute with so many zeros would likely eliminate any performance benefit of DIA/ELL.
    Columns highlighted in gray denote kernels that perform the conversion by first converting to CSR temporaries.
  }
  \centering
  {\footnotesize
  \begin{tabular}{llrrr|loo|lrr|lrr|lr|loo|lo}
    \toprule
    \multicolumn{1}{c}{\multirow{4}{*}{Matrix}} & \multicolumn{4}{c}{\smallcode{coo_csr}} & \multicolumn{3}{c}{\smallcode{coo_dia}} & \multicolumn{3}{c}{\smallcode{csr_csc}} & \multicolumn{3}{c}{\smallcode{csr_dia}} & \multicolumn{2}{c}{\smallcode{csr_ell}} & \multicolumn{3}{c}{\smallcode{csc_dia}} & \multicolumn{2}{c}{\smallcode{csc_ell}} \\
    \cmidrule(lr){2-5} \cmidrule(lr){6-8} \cmidrule(lr){9-11} \cmidrule(lr){12-14} \cmidrule(lr){15-16} \cmidrule(lr){17-19} \cmidrule(lr){20-21} 
    & {\multirow{2}{*}{\begin{tabular}{@{}c@{}}taco \\ w/ ext. \end{tabular}}} & {\multirow{2}{*}{\begin{tabular}{@{}c@{}}taco \\ w/o ext. \end{tabular}}} & \multicolumn{1}{c}{\multirow{2}{*}{skit}} & \multicolumn{1}{c|}{\multirow{2}{*}{mkl}} & \multicolumn{1}{c}{\multirow{2}{*}{\begin{tabular}{@{}c@{}}taco \\ w/ ext.\end{tabular}}} & \cellcolor{nonsymmcolor} & \cellcolor{nonsymmcolor} & \multicolumn{1}{c}{\multirow{2}{*}{\begin{tabular}{@{}c@{}}taco \\ w/ ext.\end{tabular}}} & \multicolumn{1}{c}{\multirow{2}{*}{skit}} & \multicolumn{1}{c|}{\multirow{2}{*}{mkl}} & \multicolumn{1}{c}{\multirow{2}{*}{\begin{tabular}{@{}c@{}}taco \\ w/ ext.\end{tabular}}} & \multicolumn{1}{c}{\multirow{2}{*}{skit}} & \multicolumn{1}{c|}{\multirow{2}{*}{mkl}} & \multicolumn{1}{c}{\multirow{2}{*}{\begin{tabular}{@{}c@{}}taco \\ w/ ext.\end{tabular}}} & \multicolumn{1}{c|}{\multirow{2}{*}{skit}} & \multicolumn{1}{c}{\multirow{2}{*}{\begin{tabular}{@{}c@{}}taco \\ w/ ext.\end{tabular}}} & \cellcolor{nonsymmcolor} & \cellcolor{nonsymmcolor} & \multicolumn{1}{c}{\multirow{2}{*}{\begin{tabular}{@{}c@{}}taco \\ w/ ext.\end{tabular}}} & \cellcolor{nonsymmcolor} \\
    & & & & & & \multicolumn{1}{c}{\multirow{-2}{*}{\cellcolor{nonsymmcolor} skit}} & \multicolumn{1}{c|}{\multirow{-2}{*}{\cellcolor{nonsymmcolor} mkl}} & & & & & & & & & & \multicolumn{1}{c}{\multirow{-2}{*}{\cellcolor{nonsymmcolor} skit}} & \multicolumn{1}{c|}{\multirow{-2}{*}{\cellcolor{nonsymmcolor} mkl}} & & \multicolumn{1}{c}{\multirow{-2}{*}{\cellcolor{nonsymmcolor} skit}} \\
    \midrule
pdb1HYS           & 1 (57.5) & 15.88 & 1.02 & 1.11 &          &      &      &          &      &      &          &      &      & 1 (79.1) & 1.68 &          &      &      & 1 (79.1) & 1.68 \\
jnlbrng1          & 1 (0.96) & 31.15 & 0.97 & 1.56 & 1 (0.86) & 3.07 & 3.38 &          &      &      & 1 (0.91) & 1.85 & 1.56 & 1 (0.92) & 0.95 & 1 (0.91) & 1.85 & 1.56 & 1 (0.92) & 0.95 \\
obstclae          & 1 (0.93) & 31.95 & 1.00 & 1.60 & 1 (0.86) & 3.02 & 3.36 &          &      &      & 1 (0.91) & 1.84 & 1.54 & 1 (0.82) & 1.04 & 1 (0.91) & 1.84 & 1.54 & 1 (0.82) & 1.04 \\
\rowcolor{nonsymmcolor}
chem\_master1     & 1 (1.06) & 29.05 & 1.04 & 1.44 & 1 (0.88) & 4.60 & 4.94 & 1 (1.10) & 0.98 & 2.14 & 1 (0.93) & 1.83 & 1.54 & 1 (0.91) & 0.91 & 1 (0.96) & 4.24 & 3.85 & 1 (1.24) & 3.11 \\
\rowcolor{nonsymmcolor}
rma10             & 1 (34.0) & 12.77 & 1.01 & 0.96 &          &      &      & 1 (29.1) & 1.17 & 1.16 &          &      &      & 1 (49.2) & 1.84 &          &      &      & 1 (62.8) & 2.09 \\
dixmaanl          & 1 (1.61) & 30.64 & 1.02 & 1.42 & 1 (1.50) & 5.03 & 3.11 &          &      &      & 1 (1.54) & 1.88 & 1.57 & 1 (1.35) & 0.97 & 1 (1.54) & 1.88 & 1.57 & 1 (1.35) & 0.97 \\
cant              & 1 (27.4) & 25.89 & 1.00 & 1.35 & 1 (45.3) & 4.16 & 4.39 &          &      &      & 1 (44.5) & 3.61 & 3.63 & 1 (59.8) & 1.78 & 1 (44.5) & 3.61 & 3.63 & 1 (59.8) & 1.78 \\
\rowcolor{nonsymmcolor}
shyy161           & 1 (1.69) & 26.92 & 1.00 & 1.50 & 1 (1.76) & 4.98 & 3.05 & 1 (1.64) & 1.07 & 2.36 & 1 (1.86) & 1.85 & 1.54 & 1 (1.77) & 0.94 & 1 (1.89) & 4.68 & 3.44 & 1 (2.36) & 3.00 \\
consph            & 1 (64.8) & 18.58 & 1.01 & 1.21 &   &       &       &   &       &       &   &       &       & 1 (88.9) & 1.45 &   &       &       & 1 (88.9) & 1.45 \\
denormal          & 1 (5.61) & 33.13 & 1.01 & 1.51 & 1 (5.14) & 5.10 & 5.78 &   &       &       & 1 (4.83) & 2.21 & 2.26 & 1 (5.17) & 1.02 & 1 (4.83) & 2.21 & 2.26 & 1 (5.17) & 1.02 \\
\rowcolor{nonsymmcolor}
Baumann           & 1 (4.70) & 25.21 & 0.99 & 1.49 & 1 (3.48) & 5.23 & 5.48 & 1 (4.71) & 1.03 & 1.89 & 1 (3.56) & 1.95 & 1.70 & 1 (3.33) & 1.01 & 1 (3.60) & 5.07 & 4.10 & 1 (4.74) & 3.16 \\
cop20k\_A         & 1 (63.6) &  8.46 & 0.89 & 0.96 &   &       &       &   &       &       &   &       &       & 1 (34.8) & 3.60 &   &       &       & 1 (34.8) & 3.60 \\
shipsec1          & 1 (81.6) & 18.29 & 1.02 & 1.28 &   &       &       &   &       &       &   &       &       & 1 (112)  & 1.93 &   &       &       & 1 (112)  & 1.93 \\
\rowcolor{nonsymmcolor}
majorbasis        & 1 (12.3) & 24.05 & 1.00 & 1.33 & 1 (10.9) & 3.34 & 3.70 & 1 (12.1) & 0.99 & 1.78 & 1 (9.91) & 2.43 & 2.42 & 1 (8.17) & 1.03 & 1 (10.4) & 3.47 & 4.37 & 1 (20.1) & 1.88 \\
\rowcolor{nonsymmcolor}
scircuit          & 1 (15.8) & 11.54 & 1.00 & 1.10 &   &       &       & 1 (16.4) & 0.95 & 1.09 &   &       &       &   &       &   &       &       &   &       \\
\rowcolor{nonsymmcolor}
mac\_econ\_fwd500 & 1 (11.1) & 20.52 & 0.99 & 1.29 &   &       &       & 1 (11.6) & 1.00 & 1.38 &   &       &       &   &       &   &       &       &   &       \\
pwtk              & 1 (121)  & 19.77 & 1.00 & 1.29 &   &       &       &   &       &       &   &       &       & 1 (123) & 4.09 &   &       &       & 1 (123) & 4.09 \\
Lin               & 1 (9.88) & 30.12 & 0.98 & 1.36 & 1 (8.40) & 4.87 & 5.12 &   &       &       & 1 (8.14) & 1.92 & 1.70 & 1 (10.1) & 0.98 & 1 (8.14) & 1.92 & 1.70 & 1 (10.1) & 0.98 \\
ecology1          & 1 (42.3) & 19.82 & 0.99 & 1.41 & 1 (36.8) & 2.74 & 3.00 &   &       &       & 1 (35.8) & 1.64 & 1.44 & 1 (37.5) & 1.08 & 1 (35.8) & 1.64 & 1.44 & 1 (37.5) & 1.08 \\
\rowcolor{nonsymmcolor}
webbase-1M        & 1 (57.9) & 10.27 & 1.01 & 0.99 &   &       &       & 1 (59.3) & 1.00 & 1.14 &   &       &       &   &       &   &       &       &   &       \\
\rowcolor{nonsymmcolor}
atmosmodd         & 1 (113) & 15.15 & 0.96 & 1.21 & 1 (64.9) & 3.26 & 3.49 & 1 (113) & 0.97 & 1.04 & 1 (62.2) & 1.72 & 1.58 & 1 (74.1) & 1.17 & 1 (62.9) & 3.40 & 3.39 & 1 (88.7) & 2.20 \\
    \midrule
\textbf{Geomean} & \textbf{1} & \textbf{20.39} & \textbf{1.00} & \textbf{1.29} & \textbf{1} & \textbf{4.01} & \textbf{3.96} & \textbf{1} & \textbf{1.02} & \textbf{1.48} & \textbf{1} & \textbf{2.01} & \textbf{1.80} & \textbf{1} & \textbf{1.36} & \textbf{1} & \textbf{2.75} & \textbf{2.51} & \textbf{1} & \textbf{1.78} \\
    \bottomrule
  \end{tabular}
  }
  \label{tab:matrix-results}
\end{table*}
}

We measure the performance of sparse tensor conversion routines that our technique generates for seven distinct combinations of source and target formats:
\vspace{-0.8em}
\begin{multicols}{2}
\begin{itemize}[leftmargin=1.25\parindent]
  \item COO to CSR (\code{coo_csr})
  \item COO to DIA (\code{coo_dia})
  \item CSR to CSC (\code{csr_csc})
  \item CSR to DIA (\code{csr_dia})
  \item CSR to ELL (\code{csr_ell})
  \item CSC to DIA (\code{csc_dia})
  \item CSC to ELL (\code{csc_ell})
\end{itemize}
\end{multicols}
\vspace{-0.8em}
\noindent
where inputs and outputs in COO, CSR, or CSC are not assumed to be sorted (though nonzeros are still necessarily grouped by row/column in CSR/CSC).
For each combination of formats, we also measure the performance of conversion between those formats using SPARSKIT and Intel MKL.
Both libraries implement routines that directly convert matrices from COO to CSR, CSR to CSC, and CSR to DIA.
Additionally, SPARSKIT supports directly converting matrices from CSR to ELL.
However, neither SPARSKIT nor Intel MKL implements routines that directly convert matrices between the remaining combinations of formats.
Thus, to perform those conversions using either library, we first convert the input from its source format to a temporary in CSR and then convert the temporary from CSR to the intended target format.
(If the input matrix is symmetric though, then conversions from CSC to DIA/ELL are cast as conversions from CSR to DIA/ELL, since CSR and CSC are equivalent in that case.)


\tabref{matrix-results} show the results of our experiments.
As these results demonstrate, our technique outperforms or is comparable to SPARSKIT and Intel MKL on average for all combinations of source and target formats that we evaluate.
On the whole, code that our technique emits to convert matrices from COO to CSR (\code{coo_csr}) and from CSR to CSC (\code{csr_csc}) exhibit similar performance as hand-implemented routines in SPARSKIT and somewhat better performance than Intel MKL.
This is unsurprising since our technique generates code that implement the same algorithms (variations of Gustavson's HALFPERM algorithm~\cite{Gustavson1978}) as SPARSKIT.
Our technique also emits code to perform CSR to DIA conversion that is 2.01$\times$ faster than SPARSKIT and 1.80$\times$ faster than Intel MKL on average.
SPARSKIT's implementation of \code{csr_dia} supports extracting a bounded number of nonzero diagonals from the input matrix and storing them in the output.
However, it implements this capability using an inefficient algorithm to identify the densest diagonals, thus leading to the slowdown.
In addition, \code{csr_ell} code that our technique emits is 1.36$\times$ faster than SPARSKIT on average.
This is because our technique emits code that invokes \code{calloc} to both allocate and zero-initialize the output arrays.
SPARSKIT, by contrast, takes user-allocated output arrays as arguments and separately initializes those arrays.
Furthermore, for conversions to DIA and ELL, code that our technique emits achieve even greater speedups---between 1.78 and 4.01$\times$---over SPARSKIT and Intel MKL when converting from COO or CSC.
This is because when the input matrix is nonsymmetric or stored in COO, both libraries must incur additional memory accesses to construct and then iterate over temporary CSR matrices.


Finally, we measure and compare against the performance of taco without our extensions for COO to CSR conversion.
By expressing COO to CSR conversion in index notation as tensor assignment (i.e., $A_{ij} = B_{ij}$, where $A$ and $B$ are CSR and COO matrices respectively), the techniques of \citeauthor{kjolstad2017} and \citeauthor{chou2018}~\cite{kjolstad2017,chou2018,kjolstad2019} can also generate code that performs the conversion.
As \tabref{matrix-results} also shows though, our technique emits code to perform COO to CSR conversion that is 20.4$\times$ faster on average.
The techniques of \citeauthor{kjolstad2017} and \citeauthor{chou2018} cannot reason about generating code that inserts nonzeros into CSR data structures out of order.
Thus, it must sort the input before performing the actual conversion, incurring significant additional overhead.
Furthermore, while sparse matrix formats like DIA or ELL can be cast as 3rd-order tensor formats, index notation (as described in \cite{kjolstad2017,chou2018,kjolstad2019}) cannot express assignment of a matrix to a 3rd-order tensor in a way that does not duplicate nonzeros along the extra dimension.
So without the extensions described in this paper, taco, which takes index notation as input, cannot emit code to perform end-to-end conversion for formats that store nonzeros in non-lexicographic coordinate order.

%% file: sections/related-works.tex
There is a long line of works~\cite{hicoo,Xie2018,bulucc2009,smith2015tensor,csr5,Bell2009,Ashari2014,Saad1989,Baskaran2012,bcsr,fcoo} on developing new sparse tensor formats to accelerate SpMV, sparse matrix-dense matrix multiplication (SpDM/SpMM), matricized tensor times Khatri-Rao products (MTTKRP), and other tensor computations.
These formats organize nonzeros in disparate ways to reduce memory footprint, improve cache utilization, expose parallelization opportunities, and better exploit hardware capabilities such as SIMD vector units for performance. 
All the aforementioned works rely on hand-implemented routines for converting tensors from a standard representation (e.g., COO) to their proposed formats.
These routines are often more complex than code to perform the computations that the proposed formats are designed to accelerate.
In addition, many techniques~\cite{Gustavson1978,Wang2016,GonzalezMesa2013,wengparallel2013,wengdesigning2013,cameron1993} have been proposed for accelerating transpositions of CSR matrices, which is equivalent to CSR to CSC conversion.

Existing sparse tensor and linear algebra compilers cannot generate efficient code to convert tensors between arbitrary, disparate formats.
The taco compiler~\cite{kjolstad2017,chou2018,kjolstad2019} can emit code to convert tensors between formats that store nonzeros in lexicographic coordinate order, but cannot generate complete conversion routines for structured formats like DIA and ELL.
Without the extensions described in this paper, taco also cannot emit code that computes and uses statistics about the input tensor to coordinate efficient assembly of the output tensor.
LL~\cite{ll,Arnold:2010:SVS:1863543.1863581} is a functional language that lets users define sparse matrix formats as nestings of lists and pairs that encode nonzeros in a dense matrix.
From these specifications, the LL compiler can emit code that convert dense matrices to different sparse matrix formats, but not efficient code that can directly convert between sparse matrix formats.
In the context of inspector-executor approaches for sparse linear algebra, \citeauthor{Nandy2018AbstractionsFS}~\cite{Nandy2018AbstractionsFS} build on CHiLL~\cite{chill} and SPF~\cite{STROUT201632} to show how inspectors that convert input matrices between different sparse matrix formats can be generated.
Their approach, however, requires specifications to be provided for every combination of potential source and target formats, since each specification is hard-coded to data structures used by the source and target formats.
SIPR~\cite{sipr} and techniques that \citeauthor{dutch1} proposed~\shortcite{dutch1,dutch2} can also generate sparse linear algebra code that, as sub-operations, convert matrices between different formats.
However, conversions in SIPR-generated code are performed by invoking hand-implemented operations that are hard-coded to specific source and target formats.
Meanwhile, the techniques proposed by \citeauthor{dutch1} only support a fixed set of standard sparse matrix formats.
Bernoulli~\cite{kotlyar-thesis,stodghill-thesis,bernoulli} uses a black-box protocol that provides an abstract interface for describing how sparse matrices stored in different data structures can be efficiently accessed.
However, the interface does not support assembly, so Bernoulli cannot generate code that construct sparse matrix results. 

There also exists a separate line of works~\cite{Chen2018,Neumann2011,svm,hique} on generating efficient code for query languages like SQL, which our attribute query language resembles.
(Attribute queries are analogous to GROUP BY queries on a table that stores the coordinates of nonzeros.)
In particular, HorseIR~\cite{Chen2018} lowers SQL queries to an array-based intermediate representation that is then optimized and compiled to efficient code. 
EmptyHeaded~\cite{Aberger2017} is a graph processing framework that generates efficient code to compute graph queries expressed in a Datalog-like language.
Furthermore, our approach to optimizing attribute queries is reminiscent of query rewriting systems in certain relational database systems like Starburst~\cite{Pirahesh1997,Pirahesh1992}.
All these techniques are designed for queries that may perform complex joins and aggregate data of arbitrary types.
By contrast, attribute queries are limited to aggregating tensor coordinates, which are integers.
This lets our technique lower and optimize attribute queries in ways that would be invalid for aggregations over other data types.

%% file: sections/conclusions.tex
We have described a technique for generating sparse tensor conversion routines that efficiently convert tensors between a wide range of formats.
Our technique is extensible, so users can easily add support for new source and target formats by simply specifying how to construct and iterate over tensors in those new formats.
By making it easy to work with the same data in multiple formats, each suited to a different stage of an application, our technique can greatly reduce the effort needed to optimize sparse tensor algebra applications.

That said, our technique can be further generalized in various ways to support conversions between an even wider range of formats.
A limitation of our technique is that it only supports tensor formats that can be expressed in the coordinate hierarchy abstraction we proposed in \cite{chou2018}.
This, for instance, precludes support for conversions to hybrid formats like HYB~\cite{bell2008} that decompose a tensor into multiple subtensors and store each using a different format, which the abstraction does not support. 
Coordinate remapping notation as we proposed also does not support grouping nonzeros based on statistics of the input tensor.
Such a capability could again, for instance, be useful for supporting conversions to hybrid tensor formats, which require determining how to divide up nonzeros amongst different subtensors.
Furthermore, our technique as described does not support generating fully parallelized code for CPUs or GPUs.
We believe addressing such limitations would constitute valuable future work.